\documentclass[aps,prl,floatfix,nofootinbib,superscriptaddress,reprint]{revtex4-2}
\usepackage{graphicx}
\usepackage[T1]{fontenc}
\usepackage[utf8]{inputenc}
\usepackage[UKenglish]{babel}
\usepackage{placeins}
\usepackage[binary-units]{siunitx}
\usepackage{color}
\usepackage[dvipsnames]{xcolor}
\usepackage{nicefrac}
\usepackage{amsmath}
\usepackage{amssymb}
\usepackage{mathtools}
\usepackage{bbold}
\usepackage{braket}
\usepackage[ruled,lined]{algorithm2e}
\usepackage{changes}
\usepackage{nameref,zref-xr,zref-user}
\usepackage[
pdftitle={Realistic First-Principle Mobility simulations in Organic Semiconductors},
pdfauthor={J. Ostmeyer, T. Nematiaram, P. Buividovich, A. Troisi},
bookmarks]{hyperref}
\usepackage{cleveref}

\graphicspath{{figures/},{data/}}

\newcommand{\im}{i}

\DeclareMathOperator{\tr}{Tr}

\DeclareMathOperator{\ord}{\mathcal{O}}

\newcommand{\lr}[1]{\left( #1 \right)}
\newcommand{\lrs}[1]{\left[ #1 \right]}
\newcommand{\vev}[1]{\left\langle #1 \right\rangle}
\newcommand{\expa}[1]{\exp\left( #1 \right)}

\newcommand{\pdagger}{{\phantom{\dagger}}}

\newcommand{\del}[2]{\ensuremath{\frac{\partial #1}{\partial#2}}}
\newcommand{\eto}[1]{\ensuremath{e^{#1}}}

\newcommand{\mD}{\ensuremath{\mathcal{D}}}

\newcommand{\ordnung}[1]{\ensuremath{\ord\left(#1\right)}}
\newcommand{\erwartung}[1]{\ensuremath{\left\langle#1\right\rangle}}

\newcommand{\hop}{J}
\newcommand{\dt}{{\delta \tau}}
\newcommand{\Nt}{{N_{\tau}}}
\newcommand{\HH}{\hat{\mathcal{H}}}

\providecommand{\ket}[1]{ \, | #1 \rangle }
\providecommand{\bra}[1]{ \langle #1 | \, }

\newcommand{\mycomment}[1]{}

\newcommand{\manuscript}{main manuscript}
\newcommand{\supl}{Supplementary Information}

\newcommand{\liverpool}{Department of Mathematical Sciences,
		University of Liverpool, UK
}

\newcommand{\chemistry}{Department of Chemistry,
	University of Liverpool, UK
}

\newcommand{\strathclyde}{Department of Pure and Applied Chemistry, University of Strathclyde, UK}

\begin{document}
\sloppy

	\title{First-principle quantum Monte-Carlo study of charge carrier mobility in organic molecular semiconductors}
	
	\author{Johann Ostmeyer}
	\affiliation{\liverpool}
    \author{Tahereh Nematiaram}
	\affiliation{\strathclyde}
	\author{Alessandro Troisi}
	\affiliation{\chemistry}
    \author{Pavel Buividovich}
	\affiliation{\liverpool}
	\date{\today}

	\begin{abstract}
		We present a first-principle numerical study of charge transport in a realistic two-dimensional tight-binding model of organic molecular semiconductors. We use the Hybrid Monte Carlo (HMC) algorithm to simulate the full quantum dynamics of phonons and either a single or multiple charge carriers without any tunable parameters.	We introduce a number of algorithmic improvements, including efficient Metropolis updates for phonon fields based on analytic insights, which lead to negligible autocorrelation times and allow to reach sub-permille precisions at small computational cost of $\ordnung{1\,\text{CPU-hour}}$. Our simulations produce charge mobility estimates that are in good agreement with experiment and that also justify the phenomenological Transient Localisation approach.
 \end{abstract}

	\maketitle
	
	\allowdisplaybreaks[1]
	\unitlength = 1em

\textbf{Introduction.} Organic molecular semiconductors are technologically interesting materials with a distinctive charge transport physics \cite{FratiniNikolkaNatMat2020}. The conduction and valence bands are narrow and can be described as being formed by the highest occupied and lowest unoccupied molecular orbitals of the constituent molecules. The transfer integrals between these localised orbitals undergo large thermal fluctuations, similar in magnitude to the transfer integrals themselves. With characteristic optical phonon frequencies being considerably smaller than both the temperature and the bandwidth, the resulting dynamic disorder has been extensively studied using semi-classical approaches to nuclear degrees of freedom such as Ehrenfest dynamics \cite{TroisiOrlandiPRL2006} or surface hopping \cite{GianniniCarofNatCommun2019, Giannini:2303.13163,GianniniAccChemRes2022,runeson2024chargetransportorganicsemiconductors}. These studies led to a conjecture of an unconventional Transient Localization (TL) charge transport mechanism \cite{Fratini:1505.02686,Fratini:2312.03840}, driven by transitions between electron states that are transiently localized (in the Anderson sense) at time scales shorter than inverse characteristic phonon frequencies~\cite{Ciuchi:1010.2893,Ciuchi:1210.1673,TroisiOrlandiJPC,HutschNPJComputMater2022,NematiaramADFM2020,TroisiJChemPhys2020,PackwoodJChemPhysFeynmanDiagrams2015}.

However, it is well known that semi-classical approximations for nuclear degrees of freedom (or, equivalently, phonons) may lead to incorrect conclusions about the charge transport in case of strong electron-phonon interactions \cite{RevModPhys.63.63,PhysRevB.56.4513}. In order to ensure that the popular TL scenario is not an artefact of semi-classical approximations, it is therefore extremely important to verify it in first-principle simulations without any assumptions about the classical behaviour of phonons or nuclei.

\begin{figure}[h!tpb]
	\centering
	\includegraphics[width=0.45\textwidth]{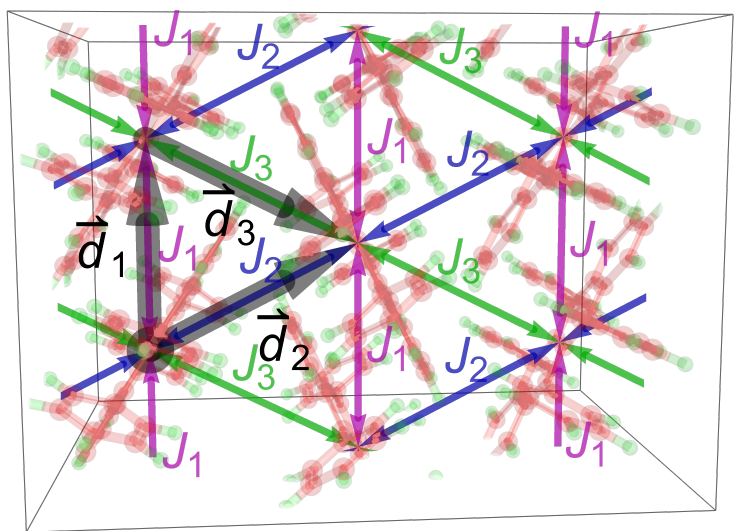}\\
	\caption{Triangular lattice structure of a rubrene crystal, a representative organic molecular semiconductor, in the high mobility plane with transfer integrals $\hop_a$ and lattice vectors $\vec{d}_a$ for each lattice bond.}
	\label{fig:rubrene-model}
\end{figure}

So far, first-principle numerical studies of charge transport in molecular semiconductors were carried out only for one-dimensional models, mostly at low charge density \cite{Mishchenko:1408.5586,deCandia:1901.03223,Wang:2203.12480,Shuai:2012.09509} (see \cite{OberhoferChemRev2017} for an overview of relevant numerical methods). 
However, the dimensionality of the model has dramatic effects on the physics involving disorder-driven localization in the electronic Hamiltonian. The vast majority of molecular semiconductors have a layered structure such that the transport takes place predominantly in a 2D plane. It was also observed that a fairly general model that captures almost the entire class of crystalline materials is formed by molecules on a triangular lattice interacting with their 3 non-equivalent neighbours through a transfer integral that is (heavily) modulated by phonons, see Fig.~\ref{fig:rubrene-model}. This class of materials is well described by the TL model in the relaxation time approximation (RTA) \cite{TroisiNatureMaterials2017}, also in the presence of high-frequency phonons that effectively renormalize the bandwidth \cite{HutschNPJComputMater2022}. However, the corresponding relaxation time $\tau_{in}$ is a phenomenological parameter that is not directly connected to the model Hamiltonian and cannot be assumed universal. 

In this paper, we put forward the Hybrid Monte Carlo (HMC) simulations of the two-dimensional tight-binding model on a triangular lattice as a first-principle reference methodology to describe charge transport in molecular semiconductors. We calculate the charge carrier mobility and its temperature dependence entirely from first principles for a broad range of model parameters, without possible artefacts of considering the nuclei classically \cite{Giannini:2303.13163,TroisiOrlandiPRL2006,GianniniCarofNatCommun2019, Giannini:2303.13163,GianniniAccChemRes2022}, introducing phenomenological quantities such as the relaxation time \cite{TroisiNatureMaterials2017}, or neglecting inter-electron interactions at high charge densities. The only input are the parameters of the 2D tight-binding Hamiltonian, such as inter-molecular transfer integrals, optical phonon frequencies and phonon coupling strengths, which are the direct output of the electronic structure calculations \cite{NematiaramADFM2020}.

\textbf{Tight-binding model and imaginary-time path integral formulation.} 
We follow \cite{TroisiNatureMaterials2017} and model the wide class of crystalline molecular semiconductors in terms of the two-dimensional tight-binding Hamiltonian describing charge carriers (holes) interacting with dispersionless phonons on an anisotropic triangular lattice:
\begin{eqnarray}
	\hat{\mathcal{H}} = 
	\sum_{k,a}\left(\frac{\hat{p}_{k,a}^2}{2} + \frac{\omega_0^2 \hat{x}_{k,a}^2}{2}\right)
	- \mu_e \sum_{k,s} \hat{c}_{k,s}^\dagger \hat{c}^\pdagger_{k,s} 
   - \nonumber \\ -	
   \label{eq:hamiltonian}
   \sum_{k,a,s} {\hop}_a \left(1-\lambda_a \hat{x}_{k,a}\right) \left(\hat{c}_{k,s}^\dagger \hat{c}^\pdagger_{k+\bar{a},s} + \hat{c}_{k+\bar{a},s}^\dagger \hat{c}^\pdagger_{k,s} \right)
  \equiv \\ \equiv
 \hat{\mathcal{H}}_{\text{ph}}\lr{\hat{x}, \hat{p}} +
 \sum_{k,l,s} \hat{c}_{k,s}^\dagger \, \lr{H_{kl}\lrs{\hat{x}} - \mu_e \, \delta_{kl} } \, \hat{c}^\pdagger_{l,s} \,,
 \label{eq:hamiltonian_sp}
\end{eqnarray}
where the indices $k,l$ label sites of a triangular lattice, $\hat{c}^\dagger_{k,s}$, $\hat{c}^\pdagger_{k,s}$ are the creation/annihilation operators for a charge carrier with spin $s$, $a$ enumerates three (forward) directions of lattice bonds between nearest-neighbour sites $k$ and $k+\bar{a}$, ${\hop}_a$ are the corresponding charge transfer integrals, $\lambda_a$ are the electron-phonon coupling constants, and $\mu_e$ is the chemical potential. We assume periodic boundary conditions. Phonons with frequency $\omega_0$ are associated with lattice bonds and described by a quadratic Hamiltonian $\hat{\mathcal{H}}_{\text{ph}}$ with amplitudes $\hat{x}_{k,a}$ and momenta $\hat{p}_{k,a}$. We also introduce a single-particle Hamiltonian $H_{kl}\lrs{\hat{x}}$ in the background of phonon fields $\hat x_{k,a}$. We do not consider electrostatic interactions between charge carriers which are negligible at low concentrations.

Our HMC simulations at finite temperature $T$ are based on the standard representation of the thermal partition function $\mathcal{Z} = \tr e^{-\beta \hat{\mathcal{H}}}$, $\beta \equiv T^{-1}$ in terms of the path integral $\int\mathcal{D} x_{k,a}\lr{\tau}$ over all possible configurations of phonon fields $x_{k,a}\lr{\tau}$ on a finite interval of Euclidean (imaginary) time $\tau \in \lrs{0, \beta}$ with periodic boundary conditions $x_{k,a}\lr{0} \equiv x_{k,a}\lr{\beta}$ \cite{Blankenbecler:PhysRevD.24.2278,Assaad:1708.03661}:
\begin{align}
\label{eq:path_integral}
\begin{split}
	\mathcal{Z} &= 
	\int\mathcal{D} x_{k,a}\lr{\tau}
	\expa{
		-S_\text{ph}\lrs{x\lr{\tau}}
	}\times \\ 
	&\quad \times
	\det{\lr{1 + 
			e^{\beta \mu_e} \, 
             \mathcal{T}e^{-\int_{0}^{\beta} d\tau H\lrs{x\lr{\tau}}}
	}}^{N_s}  ,
\end{split}
\end{align}
where $\mathcal{T}e^{\lr{\ldots}}$ is a time-ordered exponent. The Euclidean action $S_\text{ph}\lrs{x\lr{\tau}}$ for phonons is
\begin{equation*}
S_\text{ph}\lrs{x\lr{\tau}} = \int_0^{\beta}d\tau 
  \lr{
     \frac{1}{2}\lr{\frac{d x_{k,a}\lr{\tau}}{d \tau}}^2
     +
     \frac{\omega_0^2 x_{k,a}^2\lr{\tau}}{2}
  } .
\end{equation*}
The argument of the determinant in (\ref{eq:path_integral}) is a real-valued matrix for any chemical potential $\mu_e$, which implies that the path integral weight is real and non-negative for $N_s = 2$ spin components and any chemical potential $\mu_e$. Correspondingly, it can be used as statistical weight for Monte Carlo algorithms, which generate configurations of phonon fields $x_{i,a}\lr{\tau}$ with probability proportional to the integrand of (\ref{eq:path_integral}). The technical details on the derivations and numerical algorithms are provided in the \supl~\cite{supl}.

\textbf{Mobility calculation.} The charge mobility $\mu$ is calculated as the ratio $\mu = \sigma\lr{\omega \rightarrow 0}/n$ of the zero-frequency limit of the AC electrical conductivity $\sigma\lr{\omega}$ to charge density $n = \vev{\hat{c}_{k,s}^\dagger \hat{c}^\pdagger_{k,s}} \equiv \mathcal{Z}^{-1} \tr\lr{ e^{-\beta \hat{\mathcal{H}}} \hat{c}_{k,s}^\dagger \hat{c}^\pdagger_{k,s}}$. The conductivity $\sigma\lr{\omega}$ is defined in terms of the full Cartesian conductivity tensor as $\sigma\lr{\omega} = \lr{\sigma_{xx}\lr{\omega} + \sigma_{yy}\lr{\omega}}/2$. We follow the standard approach in Quantum Monte Carlo simulations~\cite{Mishchenko:1406.6486,Mishchenko:1408.5586,deCandia:1901.03223,Tripolt:1801.10348} and extract the conductivity from imaginary-time correlators of electric currents 
\begin{eqnarray}
\label{eq:current_current}
 G\lr{\tau} = \frac{1}{2 \mathcal{Z}} 
 \sum\limits_{\alpha = x, y}\tr\lr{ \hat{\mathcal{I}}_{\alpha} \, e^{-\tau \hat{\mathcal{H}}} \, \hat{\mathcal{I}}_{\alpha} \, e^{-\lr{\beta - \tau} \hat{\mathcal{H}}} } .
\end{eqnarray}
Here the electric current operator $\hat{\mathcal{I}}_{\alpha}$ is defined as
\begin{align*}
\hat{\mathcal{I}}_{\alpha} &= 
 \im \sum\limits_{k,s,a} \lr{d_a}_{\alpha}  \hop_a 
	\lr{1 - \lambda_a \hat{x}_{k,a}} 
 \lr{\hat{c}_{k+\bar{a},s}^\dagger \hat{c}^\pdagger_{k,s}
		- \hat{c}_{k,s}^\dagger \hat{c}^\pdagger_{k+\bar{a},s}}
\end{align*}
where $\lr{d_a}_{\alpha}$, $\alpha = x, y$ are the Cartesian vectors of the lattice bonds in direction $a$. We represent the correlator (\ref{eq:current_current}) as a convolution of single-particle fermionic Green's functions, which are averaged over phonon fields $x_{k,a}\lr{\tau}$.

The electrical conductivity $\sigma\lr{\omega}$ is finally estimated by inverting the Green-Kubo relations 
\begin{eqnarray}
\label{eq:green_kubo}
G\lr{\tau} = \int_0^{+\infty} \frac{d \omega}{\pi} \, 
\frac{\omega \, \cosh\lr{\omega \, \lr{\tau - \beta/2}}}
{\sinh\lr{\omega \beta/2}}
\, \sigma\lr{\omega}
\end{eqnarray}
using the Backus-Gilbert method \cite{Ulybyshev:1710.06675,Tripolt:1801.10348,ulybyshev_code_2017}.

\textbf{Monte Carlo algorithm.} To sample phonon fields $x_{k,a}\lr{\tau}$ with a probability distribution which corresponds to the path integral weight in (\ref{eq:path_integral}), we use the Hybrid Monte Carlo (HMC) algorithm~\cite{Duane1987,Buividovich:12:1,Buividovich:13:5} with a novel exact Fourier acceleration (EFA)\footnote{Following up on this work, we have generalised and benchmarked EFA in further detail~\cite{Ostmeyer:2024amv}.} inspired by~\cite{PhysRevB.99.035114,PhysRevD.32.2736,Scalettar:2203.01291} that relies on the analytic integration of the bosonic Molecular Dynamics. The HMC algorithm is universally applicable throughout the entire parameter space of the model (\ref{eq:hamiltonian}) and for any value of charge density. It provides reliable first-principle estimates of the mobility $\mu$ in run time that scales as $\mathcal{O}\lr{V^3} + O\lr{V^2 \, \Nt \, \log\Nt}$, where $V$ is the number of lattice sites and $\Nt$ is the number of discrete steps in $\tau$ used to calculate the time-ordered exponent in (\ref{eq:path_integral}) \cite{supl}. An implementation is publicly available~\cite{ssh_simulations}.

We find that our HMC algorithm with EFA is extremely efficient over the entire range of experimentally relevant Hamiltonian parameters and temperatures \mbox{$T = 100 \ldots 400 \, \mathrm{K}$} and does not suffer from long autocorrelation times~\cite{Scalettar:2005.09673,Assaad:1708.03661,supl}, in contrast to previous HMC studies of electron-phonon models \cite{Scalettar:2005.09673,Scalettar:2109.09206,Assaad:2102.08899, Cai:2308.06222} which focused on spontaneously ordered electron states.

\begin{figure}[t]
	\centering
	\resizebox{0.98\columnwidth}{!}{{\large\input{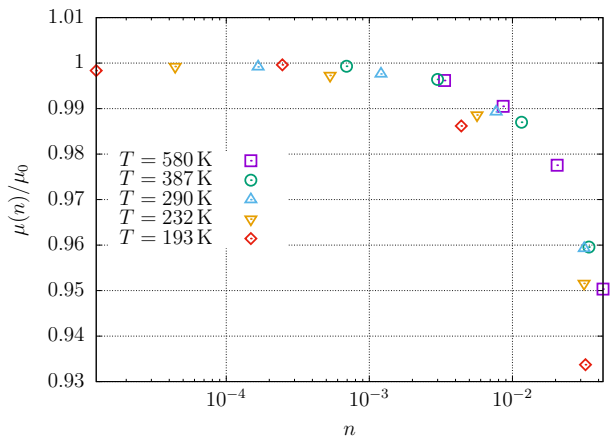}}}
	\caption{Ratio of the finite-density charge mobility $\mu\lr{n}$ to its low-density limit $\mu_0 = \mu\lr{n \rightarrow 0}$ as a function of charge density $n$ for a symmetric generic molecular semiconductor with $\hop = \SI{100}{\milli\eV}$. See \cref{fig:el_num-cond-supl} in \cite{supl} for the dependence of both quantities on the chemical potential. Errors are smaller than data points \cite{supl}.}
	\label{fig:el_num-cond}
\end{figure}

\textbf{Low-density/single charge carrier limit.} Typical charge densities in molecular semiconductors are very low, $n \lesssim 0.01$. As we explicitly demonstrate in Fig.~\ref{fig:el_num-cond}, inter-electron interactions become negligible at such low densities and it is sufficient to consider the dynamics of a single charge carrier interacting with a bath of thermal phonons \cite{OberhoferChemRev2017,Mishchenko:1408.5586,deCandia:1901.03223,Wang:2203.12480,Fratini:1505.02686}. Therefore we also consider this regime explicitly, in addition to the full finite density simulations, which allows us to accelerate our HMC algorithm even further.

Expanding the many-body partition function (\ref{eq:path_integral}) in powers of the fugacity $e^{\beta \mu_e}$, the partition function in the Hilbert subspace with charge $Q=1$ is found to be \cite{supl}
\begin{eqnarray}
\label{eq:path_int_Q1}
 \mathcal{Z}_1 = 
 \int\mathcal{D} x_{k,a}\lr{\tau}
 e^{
 -S_\text{ph}\lrs{x\lr{\tau}}
 }
 \tr  \mathcal{T}e^{-\int_{0}^{\beta} d\tau H\lrs{x\lr{\tau}}} .
\end{eqnarray}
The path integral weight in (\ref{eq:path_int_Q1}) is always positive for our range of model parameters, which allows us to use the HMC algorithm with EFA again. This provides a novel practical alternative to worldline (WLMC) \cite{deCandia:1901.03223} and Diagrammatic Monte Carlo \cite{Mishchenko:1408.5586} for polaron-type single-particle systems, which becomes particularly advantageous for charge transport in molecular semiconductors. The equivalence between HMC and the ensemble of single-particle worldlines in WLMC is readily established using the hopping expansion of the non-local factor $\tr \mathcal{T}e^{-\int_{0}^{\beta} d\tau H\lrs{x\lr{\tau}}}$. Within a few CPU-hours, the HMC yields the current-current correlator (\ref{eq:current_current}) in the single-particle Hilbert subspace on lattices with $\mathcal{O}\lr{1000}$ sites with permille-level precision, which is crucial for a numerically stable inversion of the Green-Kubo relations~\eqref{eq:green_kubo}.

\textbf{Numerical results.} To illustrate the main features of our numerical method and to compare it with the TL approach, we consider a large class of molecular semiconductors (including rubrene, see Fig.~\ref{fig:rubrene-model}) which can be mapped onto the tight-binding Hamiltonian (\ref{eq:hamiltonian}) with $\hop_2 = \hop_3 \neq \hop_1$ with sufficiently good precision \cite{TroisiNatureMaterials2017}. We use a convenient parametrization \cite{TroisiNatureMaterials2017}
\begin{eqnarray}
\label{eq:generic_model}
 \hop_1 = \hop\cos\theta, \,
 \hop_2=\hop \sin\theta/\sqrt{2}, \,
 \hop_3=\hop\sin\theta/\sqrt{2}  .
\end{eqnarray}
Thermal fluctuations of transfer integrals are usually around $\Delta \hop_a \approx \hop_a/2$ at room temperature $T_0 = \SI{25}{\milli\eV}$. From this, we estimate the electron-phonon couplings $\lambda_a$ in (\ref{eq:hamiltonian}) and assume that $\lambda_a$ are temperature-independent, so that the temperature dependence of $\Delta \hop_a$ is entirely driven by thermal phonon fluctuations $\langle\hat{x}_{k,a}^2\rangle$. When we consider the parametrization of eq.~\eqref{eq:generic_model} for a \textit{generic molecular semiconductor}, we assume a typical triangular lattice with bond lengths $|\vec{d}_a| = \SI{7.2}{\angstrom}$ (see Fig.~\ref{fig:rubrene-model}) and phonon frequencies $\omega_0 = \SI{6}{\milli\eV}$ \cite{FratiniNikolkaNatMat2020,TroisiNatureMaterials2017,supl}.

\begin{figure}[t]
	\centering
	\resizebox{0.98\columnwidth}{!}{{\large\input{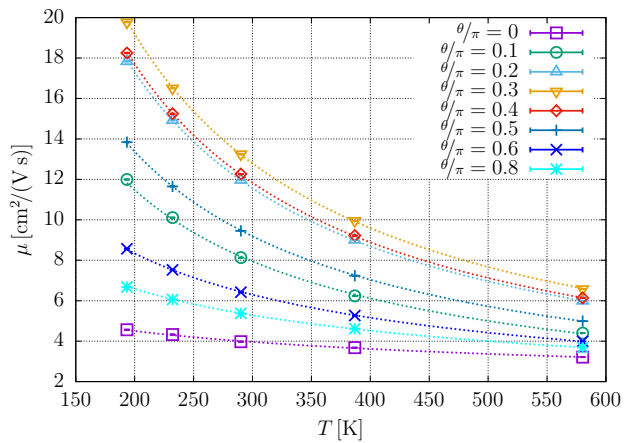}}}
	\caption{Temperature dependence of low-density charge mobility in generic molecular semiconductors with different values of $\theta$. Dashed lines are fits to the power law $\mu \sim T^{-\alpha}$ with maximal $\alpha= \num{1\pm0.005}$ for the angle region $0.2\le\nicefrac{\theta}{\pi}\le0.4$ and a monotonous decrease to $\alpha=\num{0.317\pm0.008}$ for $\theta=0$.}
	\label{fig:mobility_of_temperature}
\end{figure}

First, we check the range of applicability of the low-density/single-particle approximation by performing full HMC simulations at finite charge densities and checking how quickly the low-density limit $\mu_0 = \mu\lr{n \rightarrow 0}$ of the mobility is approached. The convergence is visualised in Fig.~\ref{fig:el_num-cond} for a symmetric generic molecular semiconductor with $\hop_1 = \hop_2 = \hop_3 = \SI{100}{\milli\eV}/\sqrt{3}$. We use lattices with $15 \times 15$ and \mbox{$18 \times 18$} lattice sites with $\Nt=48$ and $\Nt = 64$ steps for Euclidean time $\tau$ in (\ref{eq:path_integral}), thereby explicitly checking that we reach the thermodynamic and the continuum-time limits \cite{supl}. The results from \mbox{$18 \times 18$} lattices with $\Nt = 64$ shown in Fig.~\ref{fig:el_num-cond} suggest that finite-density effects become smaller than $1\%$ at $n \lesssim 0.01$ holes per lattice cell, which is in good correspondence with typical charge densities in transistor devices based on molecular semiconductors. In what follows, we will therefore present the mobility values obtained directly in the single-particle (low-density) limit from (\ref{eq:path_int_Q1}).

In Fig.~\ref{fig:mobility_of_temperature} we present the temperature dependence of the low-density charge mobility in a generic molecular semiconductor with parametrization (\ref{eq:generic_model}), fitting the data to the power law $\mu \sim T^{-\alpha}$, $\alpha > 0$ expected for band-like transport. The fits describe the data very well, supporting the experimental evidence \cite{Giannini:2303.13163} for an intermediate regime between the band-like transport and thermally activated hopping in molecular semiconductors in the experimentally relevant range of parameters and temperatures. More specifically, close to the symmetric case $\theta/\pi\approx0.3$ we find $\alpha=1$ with very high precision.

\begin{figure}[t]
	\centering
	\resizebox{0.98\columnwidth}{!}{{\large\input{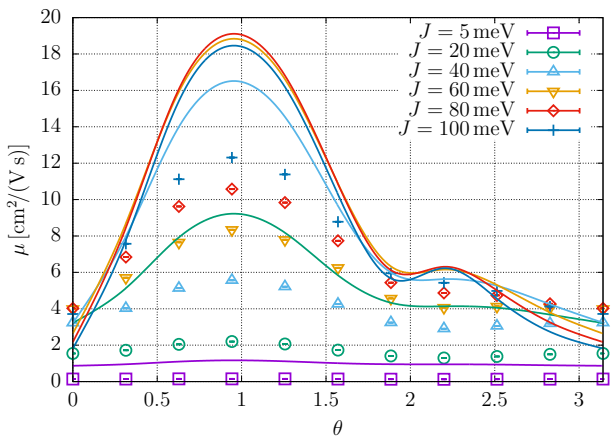}}}
	\caption{The dependence of low-density, room-temperature charge mobility $\mu$ of generic molecular semiconductors on the relative magnitudes of transfer integrals as parametrized by $\theta$ in (\ref{eq:generic_model}) from HMC (points) and TL (lines) calculations.}
	\label{fig:mobility_of_angle-combined}
\end{figure}

As a benchmark of the TL model, in Fig.~\ref{fig:mobility_of_angle-combined} we compare HMC results for the dependence of charge mobility on the relative magnitude of transfer integrals as parametrized by $\theta$ in (\ref{eq:generic_model}) with the predictions of TL RTA calculations \cite{TroisiNatureMaterials2017} (with relaxation time $\tau_{in} = 1/\omega_0$ which is typical for this method, but might not be the optimal value~\cite{PhysRevResearch.2.013001}). Both methods produce a qualitatively similar dependence on $\theta$, but the HMC yields a considerably stronger dependence on the overall hopping strength $\hop$ towards $\hop \sim \SI{100}{\milli\eV}$. Both HMC and TL yield maximal mobility for the maximally symmetric case with $\hop_1 = \hop_2 = \hop_3$, and minimal for nearly one-dimensional materials with $\hop_2 = \hop_3 = 0$. 

\begin{figure}[t]
	\centering
	\resizebox{0.98\columnwidth}{!}{{\large\input{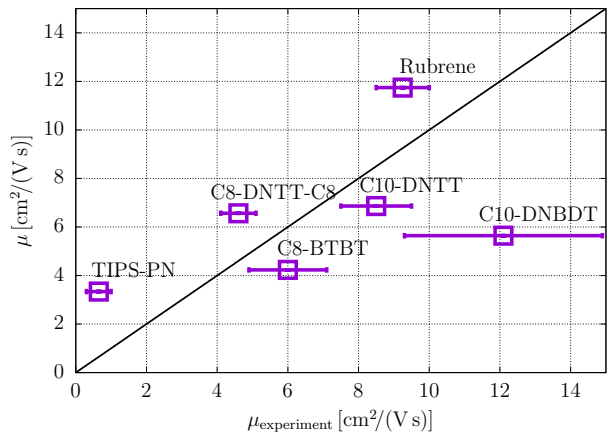}}}
	\caption{HMC results for room-temperature charge mobility vs.\ experimental results (from~\cite{TroisiNatureMaterials2017,VongJPhysChemLett2022,Giannini:2303.13163}) for six representative materials. See \supl\ \cite{supl} for details.}
	\label{fig:real-mobility_not_tl}
\end{figure}

Finally, Fig.~\ref{fig:real-mobility_not_tl} shows a comparison of our HMC results with experimental data on charge mobilities for six representative molecular semiconductors (TIPS-Pentacene, C8-DNTT-C8, C8-BTBT, C10-DNTT, Rubrene, C10-DNBDT) using exact data for the parameters of the Hamiltonian (\ref{eq:hamiltonian}) \cite{supl}. Transfer integrals $\hop_a$ with their characteristic fluctuations $\Delta \hop_a$ are obtained from ab-initio electronic structure calculations \cite{VongJPhysChemLett2022,TroisiNatureMaterials2017,Giannini:2303.13163} and lattice structures are taken from the Cambridge Structural Database \cite{CambridgeStructuralDatabase} as well as \cite{Northrup:2011,Ishii:2020,Mannsfeld:2011}. The agreement with experimental data is rather good. Since our methodology is free of any approximations given the initial Hamiltonian, any deviations from experiment are either due to experimental uncertainties or limitations of our model Hamiltonian (\ref{eq:hamiltonian}) (such as a simplified phonon spectrum with just a single frequency).

\textbf{Conclusions and outlook.} We advocate Hybrid Monte Carlo (HMC) simulations as an efficient tool for first-principle studies of charge transport in molecular semiconductors. Its computational efficiency makes the HMC well-suited for the discovery of new high-mobility semiconductors by means of large-scale scans through the space of Hamiltonian parameters \cite{NematiaramADFM2020,TroisiNatureMaterials2017}. Our results also provide a rigorous justification of the TL model without invoking any phenomenological parameters such as relaxation time. 

Our numerical approach is ready to be extended towards more realistic simulations. Since the computational complexity of the method is dominated by manipulations with fermionic matrices, additional phonon modes, nontrivial phonon dispersions, and non-linear terms in the electron-phonon and phonon-phonon interactions can all be introduced practically at no extra cost. Extrinsic disorder, which is important for the realistic description of charge localization at low temperatures $T \lesssim 150 \, K$ \cite{deCandia:1901.03223,Wang:2203.12480}, can be easily introduced as a random site-dependent chemical potential in the Hamiltonian (\ref{eq:hamiltonian}). Likewise, charge carrier interactions in doped organic semiconductors \cite{DopingOrganicSemiconductors2022} can be treated by the HMC algorithm without any approximations. The method can also easily accommodate the effects of an external magnetic field, such as the Hall current, and spin polarisation. Optical response in the THz range is also straightforward to explore.

We thank Simone Fratini for clarifications on the TL methodology. This work was funded in part by the STFC Consolidated Grant no.\ ST/T000988/1 and by the Deutsche Forschungsgemeinschaft (DFG, German Research Foundation) as part of the CRC 1639 NuMeriQS -- project no.\ 511713970. A.T.\ thanks the European Research Council (Grant No.\ 101020369) for supporting his research. Numerical simulations were undertaken on Barkla, part of the High Performance Computing facilities at the University of Liverpool, UK.

	\bibliography{bibliography}

\begin{thebibliography}{59}
\bibitem [{sup()}]{supl}%
  \BibitemOpen
  \href@noop {} {}\bibinfo {note} {See \hyperref[tit:supl]{\supl}, which
  includes Refs.~[50-59], for further details including derivations,
  algorithms, and simulation data.}\BibitemShut {Stop}%
\end{thebibliography}
	
\appendix
\clearpage
\onecolumngrid

\section*{\supl}\label{tit:supl}

\tableofcontents

\section{Band structure in the absence of phonons}\label{sec:non-interacting}

The non-interacting electronic part of the tight-binding Hamiltonian (see eq.~\eqref{eq:hamiltonian} of the \manuscript)
\begin{eqnarray}
\hat{\mathcal{H}}_0 = \sum_{k,a,s} {\hop}_a  \left(\hat{c}_{k,s}^\dagger \hat{c}_{k+\bar{a},s} + \hat{c}_{k+\bar{a},s}^\dagger \hat{c}_{k,s} \right)
 - \mu_e \sum_{k,s} \hat{c}_{k,s}^\dagger \hat{c}_{k,s}
\end{eqnarray}
becomes diagonal in momentum space featuring a single band with the dispersion relation
\begin{eqnarray}
	E(k_1, k_2) = -2 {\hop}_1 \cos(k_1) -2{\hop}_2 \cos(k_2) - 2{\hop}_3 \cos(k_2 - k_1) - \mu_e
    = \nonumber \\= 
    -\mu_0 + ({\hop}_1+{\hop}_3) \, k_1^2 + ({\hop}_2+{\hop}_3) \, k_2^2 - 2{\hop}_3 \, k_1 \, k_2 + \ordnung{k^4},\, 
    \nonumber \\
	\mu_0 \coloneqq 2({\hop}_1+{\hop}_2+{\hop}_3) + \mu_e\,,\label{eq:define_mu_0}
\end{eqnarray}
where we use a skewed coordinate system with axes $1$ and $2$ aligned along the lattice vectors $\vec{d}_1$ and $\vec{d}_2$, which correspond to hopping amplitudes (transfer integrals) $\hop_1$ and $\hop_2$ (see Fig.~\ref{fig:rubrene-model} in \manuscript).

\begin{figure}
	\centering
	\includegraphics[width=.8\textwidth]{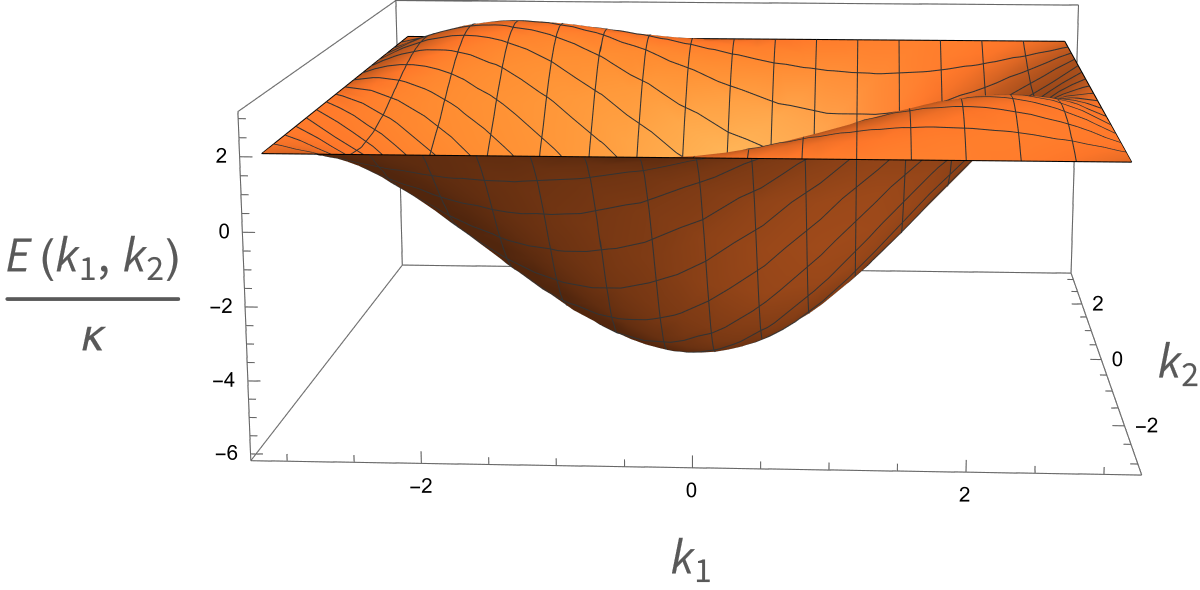}
	\caption{Single-particle dispersion relation $E(k)$ with ${\hop}_1={\hop}_2={\hop}_3={\hop}$ in the absence of phonons.}
	\label{fig:dispersion}
\end{figure}

Here, the constant $\mu_0$ can be interpreted as an effective chemical potential, directly governing the occupation of the band. $\mu_0>0$ means the Fermi surface is within the band, implying non-zero occupation and conductivity. On the other hand $\mu_0<0$ implies a gapped state in which the Fermi surface is below the band and only thermal excitations populate the band. The dispersion relation is plotted for symmetric hopping amplitudes and $\mu_0=0$ in figure~\ref{fig:dispersion}.

In what follows, we will use the generalised definition
\begin{align}
	\mu_0 &\coloneqq \mu_e + 2\sum_a \left|{\hop}_a\right|\,,
\end{align}
where the absolute value guarantees that the Fermi surface will always be below the non-interacting band if $\mu_0 < 0$, even if some of the hopping strengths ${\hop}_a$ are negative. The definition is trivially equivalent with the initial one~\eqref{eq:define_mu_0} in the case of all ${\hop}_a\ge0$. However, in general it does not preserve the convenient property that Fermi surface and lower band edge coincide at $\mu_0=0$.

\section{Physical parameters}\label{sec:parameters}

 For real materials, we give the effective values of transfer integrals after the band renormalization due to on-site optical phonons is taken into account \cite{WangPhysChemChemPhys2015BandRenormalization}.


\begin{table}[h!tpb]
	\centering
	\begin{tabular}{|l|S[separate-uncertainty=true,per-mode=symbol]|r|r|r|S[round-mode = figures,round-precision = 2]|S|l|}
		\hline
	Material & $\mu_\text{exp}~[\si{\centi\meter\squared\per\volt\per\second}]$ & $\hop_1(\Delta\hop_1/\hop_1)~[\si{\milli\eV}]$ & $\hop_2(\Delta\hop_2/\hop_2)~[\si{\milli\eV}]$ & $\hop_3(\Delta\hop_3/\hop_3)~[\si{\milli\eV}]$ & $a~[\si{\angstrom}]$ & $b~[\si{\angstrom}]$ & Sources \\\hline
TIPS-PN & 0.65\pm0.35 & -0.5(0.569) & 5.2(3.250) & 6.2(5.070) & 7.7 & 15.6 & \cite{TroisiNatureMaterials2017,VongJPhysChemLett2022,Mannsfeld:2011} \\\hline
C8-DNTT-C8 & 4.6\pm0.5 & 55.4(0.153) & 33.1(0.549) & 33.1(0.549) & 6 & 7.9 & \cite{Giannini:2303.13163} \\\hline
C8-BTBT & 6\pm1.1 & 41.4(0.297) & -29.6(0.866) & -29.6(0.866) & 6.1 & 7.4 & \cite{VongJPhysChemLett2022} \\\hline
C10-DNTT & 8.5\pm1 & 65.9(0.210) & -37.5(0.647) & -37.5(0.647) & 6 & 7.6 & \cite{TroisiNatureMaterials2017,Northrup:2011} \\\hline
Rubrene & 9.25\pm0.75 & 96.1(0.246) & 14.7(0.421) & 14.7(0.421) & 7.2 & 14.3 & \cite{TroisiNatureMaterials2017,VongJPhysChemLett2022} \\\hline
C10-DNBDT & 12.1\pm2.8 & 46.1(0.380) & 35.5(0.592) & 35.5(0.592) & 6.1 & 7.8 & \cite{TroisiNatureMaterials2017,Ishii:2020} \\\hline
``Generic'' & & $\hop\cos\theta(0.5)$ & $\hop/\sqrt{2} \sin\theta(0.5)$ & $\hop/\sqrt{2} \sin\theta(0.5)$ & 7.2 & 12.5 & \cite{TroisiNatureMaterials2017}\\\hline
	\end{tabular}
	\caption{Parameters for six representative molecular semiconductors, and a parametrization for a generic case with two equal transfer integrals (hopping amplitudes).
	$\mu_\text{exp}$ denotes experimental values of the mobility.
	The remaining parameters are used as input for the simulations of the tight-binding Hamiltonian (Eq.~\eqref{eq:hamiltonian} in the \manuscript):
	transfer integrals $\hop_a$ (rescaled by the band renormalisation factor $f$~\cite{WangPhysChemChemPhys2015BandRenormalization,VongJPhysChemLett2022} when known, or the representative value $f=\num{0.6}$ otherwise),
	their typical fluctuations $\Delta \hop_a$ so that $\lambda_a = \frac{\Delta \hop_a}{\hop_a}\vev{\hat{x}_{k,a}^2}^{-1/2}$ (see eqs.~(\ref{eq:mean_kappa_deviation},\ref{eq:x2_vev}))
	and the unit cell lengths in the common convention $a=\left|\vec{d}_1\right|$ and $b=\left|\vec{d_2}+\vec{d_3}\right|$.
	The phonon frequency $\omega_0=\SI{6}{\milli\eV}$ is used for all the materials~\cite{TroisiNatureMaterials2017,FratiniNikolkaNatMat2020}.}
	\label{tab:params}
\end{table}

The fundamental parameters of the tight-binding Hamiltonian $\hat{\mathcal{H}}$ as defined in the \manuscript\ are the ``bare'' transfer integrals $\hop_a$ and the corresponding hole-phonon coupling constants $\lambda_a$, $a = 1, 2, 3$. However, the output from electronic structure calculations is mostly presented in the literature \cite{VongJPhysChemLett2022,TroisiNatureMaterials2017,NematiaramADFM2020} in terms of the mean squared deviation 
\begin{eqnarray}
\label{eq:mean_kappa_deviation}
\Delta \hop_a = \hop_a \, \lambda_a \, \sqrt{\vev{\hat{x}_{k,a}^2}} ,
\end{eqnarray}
of the phonon-modulated transfer integrals at room temperature $T_0 = \SI{25}{\milli\eV} \approx \SI{300}{\kelvin}$. To obtain $\lambda_a$ from the room-temperature data for $\Delta \hop_a$, we express the thermal expectation value $\vev{\hat{x}_a^2}$ with respect to the quadratic phonon Hamiltonian $\hat{\mathcal{H}}_\text{ph}$ as
\begin{eqnarray}
\label{eq:x2_vev}
	\erwartung{\hat{x}_{k,a}^2} 
    = 
    \frac{1}{\omega_0}\left(\frac{1}{\eto{\omega_0/T_0}-1}+\frac12\right)
    = \frac{1}{\omega_0\,2\tanh\left(\frac{\omega_0}{2 T_0}\right)}\, .
\end{eqnarray}
Substituting (\ref{eq:x2_vev}) into (\ref{eq:mean_kappa_deviation}), we express $\lambda_a$ as
\begin{eqnarray}
\label{eq:lambda_def}
 \lambda_a 
 = 
 \left.\frac{1}{\sqrt{\vev{\hat{x}_{k,a}^2}}}
 \frac{\Delta \hop_a}{\hop_a}\right|_{T = T_0}
 =
 \sqrt{2 \, \omega_0\, \tanh\left(\frac{\omega_0}{2 T_0}\right)} \,
 \left. \frac{\Delta \hop_a}{\hop_a} \right|_{T = T_0}
\end{eqnarray}
After expressing the electron-phonon couplings $\lambda_a$ from room-temperature data, we assume that the  $\lambda_a$ are temperature-independent, and use the same value of $\lambda_a$ for all temperatures. Correspondingly, the magnitude of thermal fluctuations $\Delta \hop_a$ of transfer integrals is determined entirely by thermal fluctuations of phonon variables $x_{k,a}$.

\mycomment{
Unless stated otherwise explicitly, numerical experiments are conducted at room temperature with the following parameters:
\begin{align}
	\beta &= \SI{40}{\eV\tothe{-1}} & \text{(inverse temperature),}\\
	{\hop} &= \left(\SI{0.1}{\eV}, \SI{0.1}{\eV}, \SI{0.1}{\eV}\right) & \text{(hopping amplitude),}\\
	\lambda &= \frac{1}{2\sqrt{\erwartung{x^2}}} & \text{(electron-phonon coupling)}\\
	&\approx \frac{1}{2}\,\sqrt{2m\omega_0\tanh\left(\beta\frac{\omega_0}{2}\right)}\,\approx\,\SI{0.018}{\eV}\,,\label{eq:lambda_0th_order}\\
	\omega_0 &= \SI{0.006}{\eV} & \text{(phonon frequency),}\\
	A &= \SI{23.62}{\angstrom\squared} & \text{(unit cell area),}
\end{align}
}
\section{Path integral formalism}\label{sec:path-integral}

The finite-temperature expectation value of an operator $\hat{A}$ is written as
\begin{equation}
	\erwartung{\hat{A}}=\mathcal{Z}^{-1} \tr\lr{ \hat{A} \, \eto{- \beta \, \HH}} ,
\end{equation}
where we defined the inverse temperature $\beta \equiv T^{-1}$. The derivation of the path integral representation of the many-body partition function $\mathcal{Z} = \tr\lr{\eto{-\beta \, \HH}}$ proceeds by trotterization 
\begin{eqnarray}
\label{eq:partition_path_int}
 \mathcal{Z} = \mathrm{Tr}\lr{\prod_{t=0}^{\Nt-1}
 \eto{-\HH \, \dt}} .
\end{eqnarray}
The inverse temperature $\beta$ now defines the imaginary time extent of the system, which can be discretized into $\Nt$ intervals of length $\dt \equiv \beta/\Nt$. 

We now use the decomposition of the identity operator $\mathbb{1}$ in terms of the direct products $\ket{\psi, x} = \ket{\psi} \otimes \ket{x}$ of fermionic coherent states $\ket{\psi}$ \cite{MontvayMuenster} and eigenstates $\ket{x}$ of the phonon field operator $\hat{x}_{k,a}$
\begin{eqnarray}
\label{eq:identity_decomp}
\hat{\mathbb{1}} = 
 \int\!
 \mD\lrs{\bar{\psi},\psi,x}
 \eto{-\bar{\psi} \psi}
 \ket{\psi, x} \bra{\psi, x} ,
\end{eqnarray}
where the integration is performed over Grassmann-valued fermionic variables $\psi \equiv \psi_{k,s}$ and the real-valued variables $x \equiv x_{k,a}$ representing (spinless) charge carriers and phonons, respectively. As in the \manuscript, the index $k$ label lattice sites, $s$ labels charge carrier spin and $a$ labels lattice bond directions. The integration measure is defined as
\begin{eqnarray}
	\mD\lrs{\bar{\psi},\psi,x}
    \equiv 
    \lr{\prod\limits_{k,s} d\bar{\psi}_{k,s} d \psi_{k,s} }
    \lr{\prod\limits_{k,a} dx_{k,a} } .
\end{eqnarray}
Inserting the identity decompositions (\ref{eq:identity_decomp}) between each of the factors $\eto{-\HH \, \dt}$ in the trotterized representation (\ref{eq:partition_path_int}), we rewrite the partition function as an integral over all possible configurations of fermion and phonon variables $\psi_t \equiv \psi_{k,s,t}$ and $x_t \equiv x_{k,a,t}$ associated with $\Nt$ sequential ``imaginary time'' values $\tau_t = \frac{\beta \, t}{\Nt}$, $t  = 0 \ldots \Nt$:
\begin{eqnarray}
\label{eq:partition_sum_integral_1}
	\mathcal{Z} 
    = 
    \int\!\prod_{t=0}^{\Nt-1}\!
    \mD\lrs{ \bar{\psi}_t,\psi_t, x_t}
    \eto{-\bar{\psi}_{t+1}\cdot\psi_{t+1}}\erwartung{\psi_{t+1},x_{t+1}\left|\eto{-\HH \dt }\right|\psi_t,x_t}.
\end{eqnarray}
Phonon variables satisfy periodic boundary conditions with $x_{\Nt} = x_0$, while fermionic variables are anti-periodic with $\psi_{\Nt} = - \psi_0$, $\bar{\psi}_{\Nt} = - \bar{\psi}_0$.

The full Hamiltonian $\HH$ is a sum of terms $\HH_x\lrs{\hat{x}} = \sum_{k,l,s} \hat{c}_{k,s}^\dagger \, \lr{H_{kl}\lrs{\hat{x}} - \mu_e \, \delta_{kl} } \, \hat{c}_{l,s}
    - \sum\limits_{k,a} \frac{\omega_0^2 \hat{x}_{k,a}^2}{2}$ and $\HH_p\lrs{\hat{p}} = \sum\limits_{k,a} \frac{\hat{p}_{k,a}^2}{2}$ that only contain the phonon operator $\hat{x}_{k,a}$ or the corresponding conjugate momentum $\hat{p}_{k,a}$. For sufficiently small imaginary time step $\dt$, the exponent $\eto{-\HH \dt}$ can be represented as a product of exponents which only contain $x$- or $p$-dependent terms:
\begin{eqnarray}
\label{eq:trotter_trick}
	\eto{-\HH \dt }
	= \eto{-\dt \HH_x} \eto{-\HH_p \dt} + \ordnung{\dt^2}\,,
\end{eqnarray}
Taking into account that the basis states $\ket{\psi_t,x_t}$ in the path integral representation (\ref{eq:partition_sum_integral_1}) are the eigenstates of the $\hat{x}$ operator, and substituting the approximation (\ref{eq:trotter_trick}) into the matrix elements $\erwartung{\psi_{t+1},x_{t+1}\left|\eto{-\HH \dt }\right|\psi_t,x_t}$, we simplify the latter as
\begin{eqnarray}
	\bra{\psi_t}\bra{x_t} \eto{-\HH \dt } \ket{x_{t+1}} \ket{\psi_{t+1}} 
    = 
    \bra{\psi_{t+1}}
    \eto{-\HH_x\lrs{x} \dt }
    \ket{\psi_t}
    \bra{x_{t+1}} \eto{-\HH_p \dt } \ket{x_t} 
    + \ordnung{\dt^2}
    \propto \nonumber \\ \propto
    \bra{\psi_t} \eto{-\frac{1}{2\dt}\lr{x_{t+1}-x_t}^2 \,+\,
		\dt\!\sum_{k,l,s} \hat{c}_{k,s}^\dagger \, \lr{H_{kl}\lrs{x_t} - \mu_e \, \delta_{kl} } \, \hat{c}_{l,s} \,-\,
		\dt \frac{\omega_0^2 x_t^2}{2}} \ket{\psi_{t+1}} + \ordnung{\dt^2}
\end{eqnarray}

It remains to calculate the matrix elements in the basis of fermionic coherent states $\ket{\psi_t}$ out the fermionic part, for which we use the relation
\begin{equation}
	\erwartung{\psi\left|\eto{\sum_{A,B}c^\dagger_A O_{A,B} c_B^\pdagger}\right|\psi'}=\eto{\sum_{A,B} \bar{\psi}_A\left[\eto{O}\right]_{A,B}\psi'_B} ,
\end{equation}
valid for any Hermitian matrix $O_{A,B}$ with some generalized indices $A, B$.
This reduces the path integral representation~(\ref{eq:partition_sum_integral_1}) of the partition function $\mathcal{Z}$ to
\begin{align}
	\mathcal{Z} &= \int\!\prod_{t=0}^{\Nt-1}\!\mD\left[ \bar{\psi}_t,\psi_t,x_t\right]\eto{-\bar{\psi}_{t+1}\cdot\psi_{t+1}}\erwartung{\psi_{t+1}\left|\eto{-\HH[c^\dagger\!,c]\dt }\right|\psi_t}\eto{-\frac{1}{2\dt}\left(x_{t+1}-x_t\right)^2 \,-\,
		\dt \frac12 \omega_0^2 x_t^2} + \ordnung{\dt^2}\\
	&= \int\!\prod_{t=0}^{\Nt}\!\mD\left[\bar{\psi}_t,\psi_t,x_t\right]
	\eto{-\bar{\psi} M \psi -\frac{1}{2\dt}\left(x_{t+1}-x_t\right)^2 \,-\,
		\dt \frac 12 \omega_0^2 x_t^2} + \ordnung{\dt^2}\\
	&= \int\mD x\, \det\left( M\right)^{N_s}\,
	\eto{-S_\text{ph}} + \ordnung{\dt^2}\,,\label{eq:final_part_sum}\\
	M_{t,t'} &\coloneqq \delta_{t,t'} - \delta_{t,t'+1}\,M_t\,,\label{eq:define_M}\\
	M_t &\coloneqq\eto{\mu_e\dt + H\lr{x_t}\dt}\,,\label{eq:define_M_t}\\
	S_\text{ph} &\coloneqq \frac{1}{2\dt}\sum\limits_t\left(x_{t+1}-x_t\right)^2 \,+\,
	\dt \frac 12 \omega_0^2 \sum\limits_t x_t^2\,,
\end{align}
where $S_\text{ph}$ denotes the phonon action and $N_s$ is the number of spin components (distinct values of the spin label $s$). The non-trivial contribution to the fermion matrix $M$ is defined by the matrix $H(x)$ of the single-particle fermion Hamiltonian in the background of the phonon field $x \equiv x_{k,a}$ acting so that
\begin{align}
	\hat{c}^\dagger \cdot H(x) \cdot \hat{c} &= \sum_{k,a} {\hop}_a \left(1-\lambda x_{k,a}\right) \left(c_{k}^\dagger c_{k+\bar{a}}^\pdagger + h.c.\right)\,.
\end{align}
Note that the periodicity of the path integral elevates the initially first order Trotter formula~\cite{trotter_original} to the second order Verlet scheme~\cite{Verlet:1967}, explaining the overall $\ordnung{\dt^2}$ error.

In the continuum limit, the discrete values $\tau_t$ densely and uniformly cover the imaginary time interval $\tau \in \lrs{0, \beta}$, and the matrix $M_{t,t'}$ approaches the differential operator $M = \frac{\partial}{\partial \tau}  + H\lrs{x\lr{\tau}}$. A standard calculation \cite{Blankenbecler:PhysRevD.24.2278} establishes that this determinant can also be rewritten as in Eq.~\eqref{eq:hamiltonian} in the \manuscript:
\begin{eqnarray}
\label{eq:detM}
\det\left( M\right) = 
\det{\lr{1 + 
  e^{\beta \mu_e} \, U\lrs{x\lr{\tau}}\lr{0, \beta}
 }}
 =
 \expa{\tr\log\lr{1 + 
  e^{\beta \mu_e} \, U\lrs{x\lr{\tau}}\lr{0, \beta}}} ,
\end{eqnarray}
where $U\lrs{x\lr{\tau}}\lr{\tau_1, \tau_2} = \mathcal{T}e^{-\int_{\tau_1}^{\tau_2} d\tau H\lrs{x\lr{\tau}}}$ is the time-ordered exponent. Its discretized representation is 
\begin{eqnarray}
\label{eq:ordered_exponent_discretization}
   U\lrs{x\lr{\tau}}\lr{\tau_1, \tau_2}
   =
   \prod\limits_{t=t_1}^{t_2-1}
   \expa{-H\lrs{x_t} \, \dt} ,
\end{eqnarray}
where $t_1$ and $t_2$ are the serial numbers of the Euclidean time slices that correspond to continuum values of the Euclidean time $\tau_1$ and $\tau_2$.

To work out the path integral representation of the partition function $\mathcal{Z}_1$ in the Hilbert subspace with charge $Q = 1$, we note that this partition function is the first-order term in the expansion of the full partition function $\mathcal{Z} = \tr e^{-\beta \HH}$ in powers of the fugacity $e^{\beta \mu_e}$. Indeed, only the states with $Q = 1$ will contribute with the weight $e^{\beta \mu_e}$. States with $Q = 0$ will produce a contribution that does not depend on the chemical potential $\mu_e$, while states with $Q \geq 2$ will produce contributions of order  $\lr{e^{\beta \mu_e}}^2$ and larger. Correspondingly, to obtain the path integral representation of $\mathcal{Z}_1$, we use the path integral representation (\ref{eq:final_part_sum}) of the full partition function $\mathcal{Z}$, and expand $\det{\lr{1 + 
  e^{\beta \mu_e} \, U\lrs{x\lr{\tau}}\lr{0, \beta}
 }}^{N_s}$ in powers of $e^{\beta \mu_e}$. To this end, it is most convenient to represent $\det\lr{1 + e^{\beta \mu_e} U}^{N_s}$ as $\expa{N_s \, \tr\log\lr{1 + e^{\beta \mu_e} U}}$:
 \begin{eqnarray}
 \label{eq:det_expansion1}
  \expa{N_s \, \tr\log\lr{1 + e^{\beta \mu_e} U}}
  =
  1 + N_s \, \tr\log\lr{1 + e^{\beta \mu_e} U} + \frac{N_s^2}{2}\lr{\tr\log\lr{1 + e^{\beta \mu_e} U}}^2 + \ldots.
 \end{eqnarray}
 The log term can be in turn expanded as $\tr\log\lr{1 + e^{\beta \mu_e} U} = e^{\beta \mu_e} \tr\lr{U} - \frac{\lr{e^{\beta \mu_e}}^2}{2} \lr{\tr\lr{U}^2} + \ldots$. We conclude that 
 \begin{eqnarray}
 \label{eq:det_expansion2}
  \det\lr{M} = \expa{N_s \, \tr\log\lr{1 + e^{\beta \mu_e} U}}
  =
  1 + N_s \, e^{\beta \mu_e} \tr\lr{U} + \mathcal{O}\lr{\lr{e^{\beta \mu_e}}^2}.
 \end{eqnarray}
 Since there are no other $\mu_e$-dependent terms in the path integral (\ref{eq:final_part_sum}), we readily arrive at Eq.~\eqref{eq:path_int_Q1} in the \manuscript\ as the path integral representation of the partition function $\mathcal{Z}_1$, up to an overall normalization factor $N_s \, e^{\beta \mu_e}$.
 
An alternative way to think about this approach becomes clear when considering observables like the mobility of the form
\begin{align}
	\erwartung{\mathcal{A}} &= \frac{\erwartung{\mathcal{A}_0}}{\erwartung{n}}\,,
\end{align}
that is the observable $\mathcal{A}$ of interest is the ratio of the expectation values of some other observable $\mathcal{A}_0$ (in the case of the mobility, this corresponds to the conductivity) and the particle number $n$. The use of such a ratio can be interpreted as a reweighting procedure because the Monte Carlo sampling has not used the true probability distribution. This reweighting can easily be avoided by augmenting the initial probability distribution $p$ with a factor $n$
\begin{align}
	p &\mapsto \frac{n}{\erwartung{n}}\cdot p\,.
\end{align}
We denote the expectation value of an observable sampled according to this augmented distribution by $\erwartung{\cdot}_n$ and obtain
\begin{align}
	\erwartung{\mathcal{A}} &= \erwartung{\frac{\mathcal A_0}{n}}_n
\end{align}
without any need in reweighting. In practice $\mathcal A_0$ is often strongly correlated with $n$, so that their ratio has only small fluctuations and even small samples result in high precision.

Thus, we readily obtain the generalisation of the $\mathcal{Z}_1$ expansion in that even the full many-body quantum dynamics can be simulated efficiently with a probability distribution $\propto n\det M \eto{-S_\text{ph}}$ with corresponding adjustments of the Molecular Dynamics Hamiltonian and observables
\begin{align}
	\mathcal H_{MD} &\mapsto \mathcal H_{MD} -\log n\\
	&= \mathcal H_{MD} -\log\left(1-\tr M^{-1}\right)\,,\\
	\mathcal{A}_0 &\mapsto \frac{\mathcal A_0}{n}
\end{align}
and getting rid of the need to reweigh by the factor $1/\erwartung{n}$.
\section{Monte Carlo simulations}\label{sec:hmc}

The integrand of equation~\eqref{eq:final_part_sum} can be interpreted as a probability density and simulated using a Monte Carlo method.

We will also use Monte-Carlo to sample phonon fields with the weight proportional to the path integral weight in the single-particle partition function Eq.~\eqref{eq:path_int_Q1} in the \manuscript. This weight is real-valued, but is not guaranteed to be positive. However, in our simulations we found that $\tr\lr{U\lr{0, \beta}}$ is always bounded from below by some positive number. 

Both path integrals Eq.~\eqref{eq:path_integral} and Eq.~\eqref{eq:path_int_Q1} in the \manuscript\ are non-local functionals of phonon fields $x_{k,a,t}$, which are therefore most efficiently sampled using non-local updates. A natural choice to generate efficient nonlocal Metropolis updates is the Hybrid Monte Carlo (HMC) algorithm~\cite{Duane1987,Blankenbecler:PhysRevD.24.2278}. 

The algorithm generates configurations of bosonic variables $x\lr{\tau}$ (phonon variables $x_{k,a}\lr{\tau}$ in our case) with the weight $e^{-S\lrs{x\lr{\tau}}}$, where $S\lrs{x\lr{\tau}}$ is the full bosonic action after integrating out the fermions, by solving Hamiltonian equations of motion (Molecular Dynamics) in some auxiliary Monte-Carlo time $t_{MD}$, with an auxiliary Hamiltonian $\mathcal{H}_{MD} = \frac{\pi^2}{2} + S\lrs{x\lr{\tau}}$. Note that the bosonic variables depend on the (discretizes) imaginary time $\tau$, which in the context of Molecular Dynamics evolution is simply treated as some additional index, along with lattice indices $k$ and $a$. 

Molecular Dynamics evolution is interleaved by updates of auxiliary momenta $\pi$, which are sampled from a standard normal distribution. The role of these updates is similar to random force in Langevin simulations in that they ensure the ergodicity of Molecular Dynamics. At the end of each interval of Molecular Dynamics evolution, the new configuration of bosonic variables is accepted with a probability $\min(1,\eto{-\Delta \mathcal{H}_{MD}})$, where $\Delta\mathcal{H}_{MD}$ is the change in an auxiliary Hamiltonian.

\subsection{Natural units}

To get rid of spurious factors, we define
\begin{align}
	y &\coloneqq \sqrt{\omega_0} x\,,\\
	\kappa &\coloneqq \frac{\lambda}{\sqrt{\omega_0}}\,,\\
	\tilde\omega &\coloneqq \omega_0\dt\,,\\
	\tilde\mu &\coloneqq \mu\dt\,,\\
	\tilde \hop &\coloneqq \hop\dt\,.
  \label{eq:rescaling}
\end{align}

\subsection{Auxiliary Hamiltonian and Molecular Dynamics evolution - full many-body quantum dynamics}

With this formulation in natural units, we can write down the Molecular Dynamics Hamiltonian which corresponds to the path integral weight~\eqref{eq:final_part_sum} as
\begin{align}
	\mathcal{H}_{MD} &= -\log\det \hat M +\frac{1}{2\tilde \omega}\sum\limits_t\left(y_{t+1}-y_t\right)^2 +
	\frac{\tilde\omega}{2} y^2 + \frac12 \pi^2\,,\label{eq:hamilton_of_y}\\
	\hat M &\coloneqq 1 + \prod_t M_t\,, \label{eq:define_m_hat}
\end{align}
with $\hat M$ defined so that $\det M = \det \hat M$.
The $\pi$ field represents the canonical momentum of $x$. Consequently, the equations governing the Molecular Dynamics evolution are
\begin{align}
	\dot y &= \pi\,,\\
	\dot \pi_t &= -\tilde \omega y_t + \frac{1}{\tilde\omega}\left(y_{t+1} - 2y_t + y_{t-1}\right)
	+\tr\left[\hat M^{-1} \del{\hat M}{y_t}\right]\, ,\label{eq:force_exact}
\end{align}
where dots denote derivatives w.r.t.\ the auxiliary Molecular Dynamics time $t_{MD}$.

\subsection{Auxiliary Hamiltonian and Molecular Dynamics evolution - single-particle quantum dynamics}
\label{subsec:single_particle_HMC}

For simulations in the sector with charge $Q = 1$, the full action is $S\lrs{x\lr{\tau}} = S_\text{ph}\lrs{x\lr{\tau}} - \log\tr U\lr{0, \beta}$. The corresponding Molecular Dynamics Hamiltonian is 
\begin{align}
	\mathcal{H}_{MD} &= -\log\tr U\lr{0, \beta} +\frac{1}{2\tilde \omega}\sum\limits_t\left(y_{t+1}-y_t\right)^2 +
	\frac{\tilde\omega}{2} y^2 + \frac12 \pi^2\,,\label{eq:hamilton_of_y_sp}
\end{align}
Consequently, the equations governing the Molecular Dynamics evolution are
\begin{align}
\label{eq:force_exact_sp}
	\dot y &= \pi\,,\\
	\dot \pi_t &= -\tilde \omega y_t + \frac{1}{\tilde\omega}\left(y_{t+1} - 2y_t + y_{t-1}\right)
	+
    \frac{1}{\tr U\lr{0, \beta}} \, \tr\lr{\frac{\partial U\lr{0, \beta}}{\partial y_t}}
    \nonumber \\ &= 
    -\tilde \omega y_t + \frac{1}{\tilde\omega}\left(y_{t+1} - 2y_t + y_{t-1}\right)
	+
    \frac{e^{-\beta \mu_e}}{\tr U\lr{0, \beta}} \, \tr\lr{\frac{\partial \hat{M}}{\partial y_t}} \, ,
\end{align}
where we take into account that $\hat{M} = 1 + e^{\beta \mu_e} U\lr{0, \beta}$.


\subsection{Explicit linearised formulae}

Since the matrix exponentials in $M$ are extremely costly to compute exactly, we can approximate the fermion matrix as a product of linear terms
\begin{align}
	M_t &=\eto{\tilde\mu + H(y_t)}\\
	&= \prod_{k=1}^{n}\left[1+\gamma_k^{(n)}\left(\tilde\mu + H(y_t)\right)\right] + \ordnung{\dt^{n+1}}\,,\label{eq:linear_approx_Mt}
\end{align}
where the (generally complex-valued) coefficients $\gamma_k^{(n)}$ have to be chosen so that the product~\eqref{eq:linear_approx_Mt} is equal to the Taylor series of same order $n$.
A detailed derivation of the factorisation, comparisons with other methods as well as practical advice on an efficient implementation can be found in Ref.~\cite{lattice23}.

We used the order $n=4$ in all our calculations as we found it to provide a good trade off between fast computability and high accuracy.

Using the factorised approximation of order $n=1$ (higher $n$ follow directly from the product rule for derivatives) as well as $\hat M$ instead of $M$ in the force term~\eqref{eq:force_exact}, allows to simplify the trace
\begin{align}
	\tr\left[M^{-1} \del{M}{y_{ia,t}}\right] &= \tr\left[\hat M^{-1} \del{\hat M}{y_{ia,t}}\right]\\
	&= \tr\left[\hat M^{-1} \left(\prod_{s=0}^{t-1}M_s\right)\left(-\tilde{\hop}_a \kappa\delta_{\erwartung{i,i+a}}\right)\left(\prod_{s=t+1}^{\Nt-1}M_s\right)\right]\\
	&= -\tilde {\hop}_a \kappa\left[\hat e_i Q_{t+1}\hat M^{-1} P_t \hat e_{i+a} + \hat e_{i+a} Q_{t+1}\hat M^{-1} P_t \hat e_{i}\right]\\
	&= -\tilde {\hop}_a \kappa\left[\left(Q_{t+1}\hat M^{-1} P_t\right)_{i,i+a} + \left(Q_{t+1}\hat M^{-1} P_t\right)_{i+a,i}\right]\,,\label{eq:ferm_force_lin}\\
	P_t &\coloneqq \prod_{s=0}^{t-1}M_s\,,\label{eq:product_p_t}\\
	Q_{t} &\coloneqq \prod_{s=t}^{\Nt-1}M_s\,,\label{eq:product_q_t}
\end{align}
where $\prod_{s=0}^{t}M_s \equiv M_0M_1\cdots M_t$ is an ordered product and $\hat e_i$ is the $i$-th standard basis vector. Here we have contracted the trace with the nearest neighbour hopping term $\delta_{\erwartung{i,i+a}}$ explicitly. 

For the Molecular Dynamics equations (\ref{eq:force_exact_sp}), the corresponding expression simplifies to
\begin{align}
\label{eq:ferm_force_lin_sp}
	\tr \del{\hat M}{y_{ia,t}} 
    &= 
    \tr\left[\left(\prod_{s=0}^{t-1}M_s\right)\left(-\tilde{\hop}_a \kappa\delta_{\erwartung{i,i+a}}\right)\left(\prod_{s=t+1}^{\Nt-1}M_s\right)\right]
    \nonumber \\ &= 
    -\tilde {\hop}_a \kappa\left[\hat e_i Q_{t+1} P_t \hat e_{i+a} + \hat e_{i+a} Q_{t+1} P_t \hat e_{i}\right]
    =
    -\tilde {\hop}_a \kappa\left[\left(Q_{t+1} P_t\right)_{i,i+a} + \left(Q_{t+1} P_t\right)_{i+a,i}\right]\, .
\end{align}

\subsection{Runtime}

Using the linearised form~\eqref{eq:linear_approx_Mt} of $M_t$ for matrix-vector products and dense matrix algebra otherwise, the evaluation of equation~\eqref{eq:ferm_force_lin} scales as $\ordnung{V\Nt} + \ordnung{V^2}$ with the spatial volume $V$ and the number of time-slices $\Nt$. The total force vector then requires to perform this calculation $\ordnung{V\Nt}$, that is $\ordnung{V^2\Nt^2} + \ordnung{V^3\Nt}$ operations in total. The memory requirement at any given time is of $\ordnung{V\Nt} + \ordnung{V^2}$.

The scaling can be changed significantly by constructing and storing the matrix products $Q_{t+1}\hat M^{-1} P_t$ explicitly.
More specifically, we used an advanced algorithm with
\begin{align}
	\text{runtime} &= \ordnung{V^2\Nt\log \Nt} + \ordnung{V^3}\,,\\
	\text{memory} &= \ordnung{V^2\Nt}
\end{align}
proceeding as follows: start with $\hat M^{-1}$ (0th step), multiply $Q_{\left\lfloor \nicefrac{\Nt}{2}\right\rfloor} \hat M^{-1}$, $\hat M^{-1} P_{\left\lfloor \nicefrac{\Nt}{2}\right\rfloor}$ (1st step), then iteratively constructing $Q_{\left\lfloor \nicefrac{3}{4}\Nt\right\rfloor} \hat M^{-1}$, $Q_{\left\lfloor \nicefrac{\Nt}{2}\right\rfloor} \hat M^{-1} P_{\left\lfloor \nicefrac{\Nt}{4}\right\rfloor}$, $Q_{\left\lfloor \nicefrac{\Nt}{4}\right\rfloor} \hat M^{-1} P_{\left\lfloor \nicefrac{\Nt}{2}\right\rfloor}$, $\hat M^{-1} P_{\left\lfloor \nicefrac{3}{4}\Nt\right\rfloor}$ (2nd step) and so on up to the final $\left\lceil\log_2 \Nt\right\rceil$-th step.

The $\ordnung{V^3}$-term is a consequence of inverting the dense matrix $\hat M$ exactly. It proved sub-dominant for the volumes investigated, so we did not speed it up any further. This computational complexity can be reduced to sub-cubic in $V$ if one uses pseudo-fermions and iterative matrix inversion algorithms \cite{Duane1987,Assaad:1708.03661}.

\section{Fourier acceleration}\label{sec:fourier_acc}

We remark that the algorithm as described in the following was first developed and used for this work. Since then, however, we have further improved and generalised the method. These updates can be found in Ref.~\cite{Ostmeyer:2024amv} and an implementation is publicly available~\cite{ssh_simulations}.

The fermionic forces $\tr\left[\hat M^{-1} \del{\hat M}{y_t}\right]$ are very small as a rule in the regime of low charge density we are interested in. Intuitively they are close to proportional to the electron number. More rigorously, the forces scale asymptotically as $\eto{\beta\mu_e}$, thus decaying exponentially as $\mu_e \rightarrow-\infty$. In realistic simulations around 1\% filling, the magnitude of the fermionic terms (``fermionic forces'') in Molecular Dynamics equations (\ref{eq:force_exact}) and (\ref{eq:force_exact_sp}) can be easily less than $10^{-5}$ times that of the corresponding bosonic forces $-\tilde \omega y_t + \frac{1}{\tilde\omega}\left(y_{t+1} - 2y_t + y_{t-1}\right)$. Therefore, it is crucial for efficient simulations to integrate different contributions to the Molecular Dynamics equations using different time steps.

In Ref.~\cite{Scalettar:2203.01291} this scale separation is exploited by assigning different time scales to the different forces in the integrator, i.e.\ for every fermionic force update a high number of small bosonic updates are performed. In addition, Ref.~\cite{Scalettar:2203.01291} introduces a `Fourier acceleration' of the form that the momenta are not sampled identically, instead faster modes are suppressed, significantly reducing the maximal forces.

Here we introduce a novel method following the same ideas as~\cite{Scalettar:2203.01291}. We also consider bosonic and fermionic forces separately, but we integrate the bosonic part analytically in Fourier space. This is mathematically equivalent to having infinitely many bosonic steps per fermionic update, simultaneously being much more precise (only limited by machine precision) and significantly faster\footnote{Often, more than $10^3$ discrete time steps are required in order to reach a similar bosonic precision as is naturally obtained by the small fermionic forces. This can have a relevant impact on the runtime, even though a single bosonic update is very fast.} than the method proposed in~\cite{Scalettar:2203.01291}. The algorithmic details of this `exact Fourier acceleration' (EFA) will be explained below and are summarised in the algorithms~\ref{alg:efa} and~\ref{alg:leap-frog}. We remark directly that the (approximate) Fourier acceleration from~\cite{Scalettar:2203.01291} is made obsolete by our EFA as larger bosonic forces are computationally no more expensive than small ones. This difference allows for a further major advantage of the EFA, namely extremely short autocorrelation times.

The purely bosonic equations of motion read
\begin{align}
	\dot y &= \pi\,,\\
	\dot \pi_t &= -\tilde \omega y_t + \frac{1}{\tilde\omega}\left(y_{t+1} - 2y_t + y_{t-1}\right)\,.
\end{align}
Both equations are diagonal after a Fourier transformation
\begin{align}
	Z_\xi\coloneqq \mathcal{F}[z]_{\xi} &\equiv \frac{1}{\sqrt{\Nt}} \sum_{t=0}^{\Nt-1} z_t\, \eto{-\im \frac{2\pi}{\Nt} t\, \xi}
\end{align}
in time direction (all spatial coordinates are left as they are)
\begin{align}
	\dot Y &= \Pi\,,\\
	\dot \Pi_\xi &= -\frac{\tilde\omega^2 + 4\sin^2\left(\frac{\pi}{\Nt}\xi\right)}{\tilde\omega}\, Y_\xi\,,
\end{align}
defining $Y \coloneqq \mathcal{F}[y]$, $\Pi\coloneqq \mathcal{F}[\pi]$ and $\xi =0,\dots,\Nt-1$. Harmonic oscillator equations like these are well known and it can easily be checked that they are solved by
\begin{align}
	Y(h) &= \cos(w_\xi h) Y^0 + \frac{1}{w_\xi} \sin(w_\xi h) \Pi^0\,,\\
	\Pi(h) &= \cos(w_\xi h) \Pi^0 - w_\xi \sin(w_\xi h) Y^0\,,\\
	w_\xi^2 &\coloneqq \frac{\tilde\omega^2 + 4\sin^2\left(\frac{\pi}{\Nt}\xi\right)}{\tilde\omega}\,,\label{eq:phonon_frequency}
\end{align}
for any given Monte Carlo time $h$ and initial configurations $Y^0=\mathcal{F}[y^0]$, $\Pi^0=\mathcal{F}[\pi^0]$. Finally, the exact time evolved configurations are obtained by reverse Fourier transformation $y(h)=\mathcal{F}^{-1}[Y(h)]$, $\pi(h)=\mathcal{F}^{-1}[\Pi(h)]$. Using fast Fourier transforms (FFT), the whole process has a runtime of $\ordnung{V\Nt\log \Nt}$ and is therefore negligible compared to the runtime of the fermionic force calculation. The entire procedure as detailed above can be found in algorithm~\ref{alg:efa}.

\begin{algorithm}
	\caption{Single time step of the bosonic time evolution in the exact Fourier acceleration (EFA).}\label{alg:efa}
	\SetKwInOut{Input}{input}
	\SetKwInOut{Output}{output}
	\Input{initial fields $y^0$, momenta $\pi^0$, time step $h$ (and parameters $\tilde \omega$, $\Nt$)}
	\Output{final fields $y(h)$ and momenta $\pi(h)$}
	$Y^0 \gets \mathcal{F}[y^0]$\;
	$\Pi^0 \gets \mathcal{F}[\pi^0]$\;
	\For{$\xi \gets 0\dots \Nt-1$}{
		$w_\xi \gets \sqrt{\frac{\tilde\omega^2 + 4\sin^2\left(\frac{\pi}{\Nt}\xi\right)}{\tilde\omega}}$\;
		$Y_{\xi} \gets \cos(w_\xi h) Y^0_\xi + \frac{1}{w_\xi} \sin(w_\xi h) \Pi^0_\xi$\;
		$\Pi_\xi \gets \cos(w_\xi h) \Pi^0_\xi - w_\xi \sin(w_\xi h) Y^0_\xi$\;
	}
	$y(h) \gets \mathcal{F}^{-1}[Y]$\;
	$\pi(h) \gets \mathcal{F}^{-1}[\Pi]$\;
\end{algorithm}

\begin{figure}[htb]
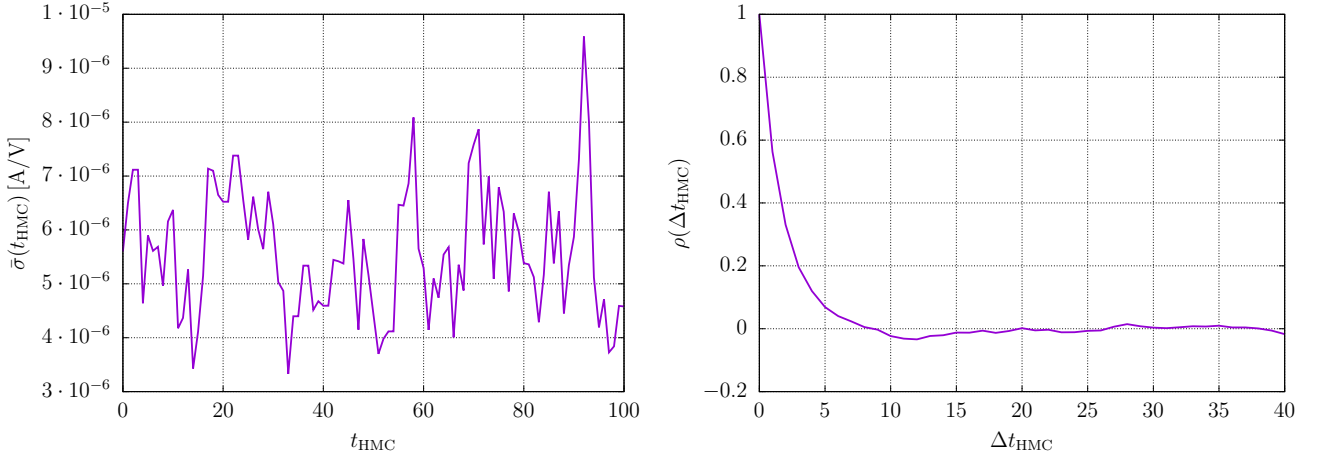

	\centering
	\resizebox{0.98\textwidth}{!}{{\large\input{data/cond_time-series}\input{data/cond_auto-corr}}}
	\caption{Time series of the conductivity $\bar \sigma$ from eq.~\eqref{eq:sigma_bar} (left) and its autocorrelation function $\rho$ (right) as a function of Monte Carlo time $t_\text{HMC}$ (one step in $t_\text{HMC}$ corresponds to one HMC trajectory) for a symmetric generic molecular semiconductor at room temperature with $\hop_1 = \hop_2 = \hop_3 = \SI{100}{\milli\eV}$. Full finite density ($n=\num{0.00775\pm.00003}\approx1\%$) simulations at reduced chemical potential $\mu_0=\SI{-50}{\milli\eV}$. Following the Ulli Wolff method~\cite{monte_carlo_errors} we find the integrated autocorrelation time in this case to be $\tau_\text{int}=\num{1.8\pm0.1}$.}
	\label{fig:time-series}
\end{figure}

Note that the slowest mode is $w_0 = \sqrt{\tilde\omega}$ and therefore a molecular dynamics trajectory length of order
\begin{align}
	h &\sim \frac{1}{\sqrt{\tilde\omega}}
\end{align}
allows to fully decouple successive configurations in the Markov chain. Throughout our simulations we used a trajectory length of exactly $\nicefrac{1}{w_0}$, typically leading to an integrated autocorrelation time of $\tau_\text{int} \approx 2$ for all measured observables, see figure~\ref{fig:time-series} for an example. Trajectories of this length usually feature an acceptance $\ge90\%$ with just one fermionic force evaluation. Any other trajectory length of the same order would be a reasonable choice too, as long as it is an irrational multiple of the period $\nicefrac{2\pi}{w_0}$.

\begin{figure*}[t]
	\centering
	\resizebox{0.98\textwidth}{!}{{\large\input{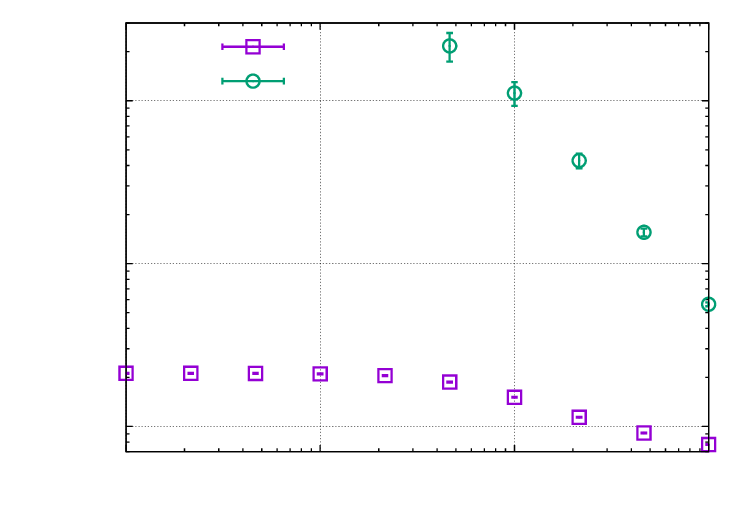}\input{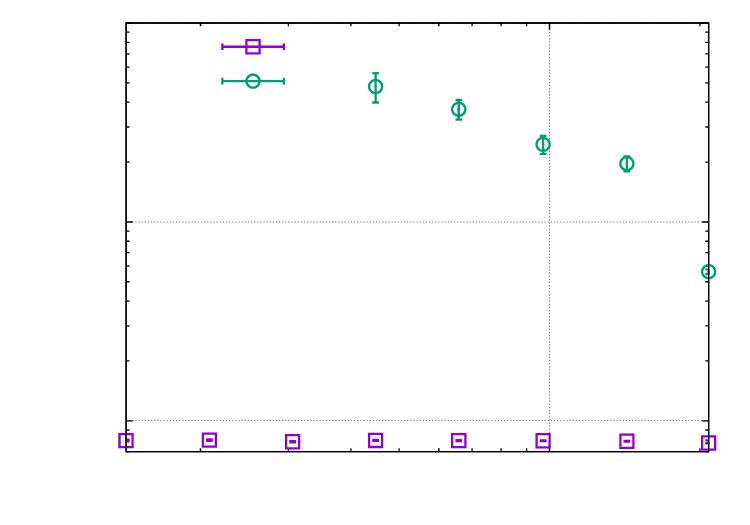}}}
	\caption{Relative error $\frac{\Delta\bar\mu}{\bar\mu}$ of the mobility $\bar\mu=\frac{\bar\sigma}{\erwartung{n_\text{el}}}$ on a $10\times 10$ lattice with $N_\tau$ imaginary time slices. The physical parameters of rubrene (see tab.~\ref{tab:params}) at room temperature $\beta=\SI{40}{\per\eV}$ with $\mu_0=0$ are used. As in the \manuscript, we used $\num{10000}$ HMC trajectories. Left: different free phonon frequencies $\omega_0$ at constant $N_\tau=48$ ($\omega_0\approx \SI{6}{\milli\eV}$ is the physical value); Right: different $N_\tau$ at fixed $\omega_0=\SI{1}{\eV}$. All simulations have similar acceptance ($\gtrsim 80\%$) and compute time per measurement. The error without EFA could not be estimated reliably for small $\omega_0$ and large $N_\tau$ with reasonable compute resources.}
	\label{fig:ssh_rubrene}
\end{figure*}

The resulting error of the mobility has been benchmarked representatively for rubrene in figure~\ref{fig:ssh_rubrene}. With EFA, it is at the permille-level, practically independent of the phonon frequency $\omega_0$ and the number of imaginary time slices $N_\tau$. Without EFA, the error increases dramatically when both parameters approach the physically interesting region. These HMC simulations were performed in the full many-body Hilbert space and the results in the single particle sector have even higher precision. 

We used the leapfrog symplectic integrator~\cite{Verlet:1967,trotter_omelyan} which is more than sufficiently precise for the physically relevant parameter set. Since the fermionic force is so small, it is highly advisable to evaluate it in the middle of a time step, see algorithm~\ref{alg:leap-frog}. This choice (so-called `position version') has roughly half the error of the alternative with a fermionic half step at either end of the trajectory~\cite{OMELYAN2003272}.

\begin{algorithm}
	\caption{Single update step with the leap-frog integrator and EFA (alg.~\ref{alg:efa}).}\label{alg:leap-frog}
	\SetKwInOut{Input}{input}
	\SetKwInOut{Output}{output}
	\Input{initial fields $y^0$, momenta $\pi^0$, time step $h$ (and matrix $\hat M$)}
	\Output{final fields $y(h)$ and momenta $\pi(h)$}
	$(y,\pi) \gets \text{EFA}\left(y^0,\pi^0,\nicefrac h2\right)$\;
	$\pi \gets \pi + h\cdot\tr\left[\hat M^{-1} \del{\hat M}{y}\right]$\;
	$\left(y(h),\pi(h)\right) \gets \text{EFA}\left(y,\pi,\nicefrac h2\right)$\;
\end{algorithm}

For simulations in the single-particle sector, the fermionic force in Algorithm \ref{alg:leap-frog} should be replaced with a corresponding expression from (\ref{eq:force_exact_sp}).

\section{Observables}\label{sec:observables}

In this work we consider expectation values and correlators of fermionic bilinear operators of the form $\hat{\mathcal{O}} = \sum\limits_{k,l,s} \hat{c}^{\dagger}_{k,s} \mathcal{O}_{k,l} \hat{c}^{\pdagger}_{l,s}$, which can be readily expressed in terms of fermionic Green's functions \cite{Blankenbecler:PhysRevD.24.2278}. For reference, in this Section we give discretized expressions for these expectation values and correlators.

Expectation value of a single fermionic bilinear operator is 
\begin{align}
	\erwartung{\hat{\mathcal{O}}} &=  \frac{1}{\Nt} \sum\limits_{t=0}^{\Nt}\erwartung{\tr\left[\hat{M}^{-1} P_t\,\mathcal{O}\,Q_t\right]}\,,\label{eq:calculate_operator}
\end{align} 
where we used the reduced fermion matrix $\hat M$ from equation~\eqref{eq:define_m_hat} as well as the discretized transfer matrices $P_t$, $Q_t$ as defined in equations~\eqref{eq:product_p_t} and \eqref{eq:product_q_t}. The averaging is performed over all configurations of phonon variables $x_{k,a,t}$ with the weight proportional to the path integral weight in the partition function~\eqref{eq:path_integral}.

Similarly, for a product of two operators, one finds 
\begin{align}
	\mathcal{Z}^{-1}
    \tr\lr{\hat{\mathcal{O}}_1 e^{-\tau \HH} \hat{\mathcal{O}}_2 e^{-\lr{\beta - \tau} \HH} } &= \erwartung{\tr\left[R_{t,0}\,\mathcal{O}_1\,T_{0,t}\,\mathcal{O}_2\right]}\,,\label{eq:calculate_2_operator_prod}\\
	R_{tt'} &\coloneqq Q_{t}\hat{M}^{-1} P_{t'}\,,\\
	T_{tt'} &\coloneqq P_{t}^{-1}\hat{M}^{-1} P_{t'} \equiv Q_{t}\hat{M}^{-1} Q_{t'}^{-1}\,,
\end{align}
where $t$ is the sequential number of discrete Euclidean time value $\tau = \frac{t \, \beta}{\Nt}$ in the correlator (\ref{eq:calculate_2_operator_prod}).

A reduction of these observables to a single-particle Hilbert space can be obtained by taking the low-density limit $\mu_e \rightarrow -\infty$. In this limit, $\hat{M} = 1 + e^{\beta \mu_e} U\lr{0, \beta}$ approaches the identity matrix, and the above expressions simplify as
\begin{align}
	\erwartung{\hat{\mathcal{O}}} &=  \frac{1}{\Nt} \sum\limits_{t=0}^{\Nt}\erwartung{\tr\left[P_t\,\mathcal{O}\,Q_t\right]}\,,\label{eq:calculate_operator_sp}
\end{align} 
\begin{align}
	\mathcal{Z}^{-1}
    \tr\lr{\hat{\mathcal{O}}_1 e^{-\tau \HH} \hat{\mathcal{O}}_2 e^{-\lr{\beta - \tau} \HH} } &= \frac{1}{{\Nt}^2} \erwartung{\tr\left[R_{t,0}\,\mathcal{O}_1\,T_{0,t}\,\mathcal{O}_2\right]}\,,\label{eq:calculate_2_operator_prod_sp}\\
	R_{tt'} &\coloneqq Q_{t} P_{t'}\,,\\
	T_{tt'} &\coloneqq P_{t}^{-1} P_{t'} \equiv Q_{t} Q_{t'}^{-1}\,.
\end{align}
Note that these observables also scale as $e^{\beta \mu_e}$, and need to be normalised by fermion number expectation value for a meaningful low-density limit. In particular, both fermion number and current-current correlators approach zero as $e^{\beta \mu_e}$ as $\mu_e \rightarrow -\infty$, but their ratio remains finite and yields charge mobility.

\subsection{Electric current operators and current-current correlators}

External electromagnetic fields are introduced in the tight-binding Hamiltonian~\eqref{eq:hamiltonian} via the Peierls substitution \cite{Peierls:33:1,Buividovich:12:1} in the hopping part of the Hamiltonian:
\begin{align}
	\HH_\text{hop} &= -\sum_{k,a,s} \chi_{k,a} \left(\eto{\im \vec{d}_a \cdot \vec{A}} \, \hat{c}_{k,s}^\dagger \hat{c}_{k+\bar{a},s}^\pdagger + \eto{-\im \vec{d}_a \cdot \vec{A}} \, \hat{c}_{k+\bar{a},s}^\dagger \hat{c}_{k,s}^\pdagger\right)\,,\\
	\chi_{k,a} &\coloneqq {\hop}_a \left(1-\lambda x_{k,a}\right) ,
\end{align}
where $\vec{A}$ is the electromagnetic vector potential and $\vec{d}_a$ is the lattice vector associated with lattice bond in direction $a$. As usual, since the experimentally relevant range of photon wavelengths is much larger than any other relevant length scale in the problem (unit cell size, lattice size, localization length), we only take into account the time dependence of $\vec{A}$ and neglect its spatial dependence. 

The current operator is given by the variation of the Hamiltonian with respect to electromagnetic vector potential:
\begin{align}
	\hat{\mathcal{I}}_{\alpha} &\equiv -\left.\del{\HH}{A_{\alpha}}\right|_{\vec{A}=0}\\
	&= \sum\limits_{k,a} \im \lr{d_a}_{\alpha} \chi_{k,a} \left(\hat{c}_{k,s}^\dagger \hat{c}_{k+\bar{a},s}^\pdagger - \hat{c}_{k+\bar{a},s}^\dagger \hat{c}_{k,s}^\pdagger\right) ,
    \label{eq:many_body_current}
\end{align} 
where $\alpha = x, y$ labels Cartesian coordinates in the high-mobility plane. From (\ref{eq:many_body_current}) we read off the corresponding single-particle current operator
\begin{eqnarray}
\label{eq:single_particle_current}
 \vec{I}_{kl} = - \im \sum\limits_{a} \vec{d}_a \lr{\chi_{k,a} \delta_{k,l-\bar{a}} - \chi_{k-\bar{a},a} \delta_{k,l+\bar{a}} } .
\end{eqnarray}
The corresponding current-current correlator 
\begin{eqnarray}
\label{eq:curr_curr_def}
   G_{\alpha\beta}\lr{\tau} = 
	\frac{1}{V \, \mathcal{Z}}
    \tr\lr{\hat{\mathcal{I}}_{\alpha} e^{-\tau \HH} \hat{\mathcal{I}}_{\beta} e^{-\lr{\beta - \tau} \HH} }
\end{eqnarray}
can be represented in terms of single-particle fermionic matrices as
\begin{align}
	G_{\alpha\beta}(\tau_t) = \\
	\begin{split}
		&= \frac{1}{V} \sum_{k,l}\left\langle
        \lr{d_a}_{\alpha} \, \lr{d_b}_{\beta} \,
        \chi_{k,a,0} \, \chi_{l,b,t} \left[
		\left(R_{t,0}\right)_{l,k}\left(T_{0,t}\right)_{k+\bar{a},l+\bar{b}}
		+\left(R_{t,0}\right)_{l+\bar{b},k+\bar{a}}\left(T_{0,t}\right)_{k,l}\right.\right.\\
		&\qquad\qquad\qquad\qquad\;\left.\left.
		-\left(R_{t,0}\right)_{l+\bar{b},k}\left(T_{0,t}\right)_{k+\bar{a},l}
		-\left(R_{t,0}\right)_{l,k+\bar{a}}\left(T_{0,t}\right)_{k,l+\bar{b}}
		\right]\right\rangle\,,
	\end{split}\label{eq:cur-cur_corr}\\
	R_{t,0} &= Q_{t}\hat{M}^{-1}\,,\\
	T_{0,t} &= \hat{M}^{-1}P_{t}\, .
\end{align}
The number of floating-point operations needed to calculate $G_{\alpha\beta}(\tau_t)$ for all discrete values $\tau_t$ of Euclidean time $\tau$ scales as $\ordnung{V^2\Nt}$.

In this work we calculate the electric conductivity and charge mobility by averaging over all spatial directions, which is equivalent to taking half of the trace of the conductivity tensor. Correspondingly, we take the trace of (\ref{eq:cur-cur_corr}) over the Cartesian indices $\alpha$ and $\beta$, leading to the current-current correlators $G\lr{\tau} = \frac{1}{2} \sum\limits_{\alpha} G_{\alpha\alpha}$ considered in the \manuscript. 

\subsection{Electrical conductivity}\label{sec:conductivity}

We calculate charge mobility from the zero-frequency limit of the optical conductivity $\sigma\lr{\omega}$. $\sigma\lr{\omega}$ is extracted from imaginary-time current-current correlators $G\lr{\tau}$ and $G_1\lr{\tau}$ by inverting the Green-Kubo relations~\eqref{eq:green_kubo}. To this end we use the Backus-Gilbert method \cite{Tripolt:1801.10348} with Tikhonov filtering in SVD decomposition (option `$+3$' for the `lambda' parameter), as implemented in~\cite{ulybyshev_code_2017}. The cut-off parameter $\lambda$ has been varied over several orders of magnitude $10^{-12}\le\lambda\le10^{-5}$ and a plateau in the value $\sigma\lr{\omega \rightarrow 0}$ has been identified around $\lambda=10^{-9}$ which has been used for the results displayed in the manuscript. As a sanity check, we also compared our results with the less precise but significantly simpler mid-point formula (see eq.~(68) in~\cite{Buividovich:12:1})
\begin{align}
	\sigma &\approx \bar \sigma \coloneqq \frac{\beta^2}{\pi} G(\beta/2)
\end{align}
and found the expected level of correspondence within ca.\ $10\%$.

Furthermore, we introduce the conversion factor $\nicefrac{\mathrm{e}^2}{\hbar}\approx\SI{2.434e-4}{\siemens}$ to SI units\footnote{Unit Siemens: $\si{\siemens}=\si{\ampere\per\volt}=\si{\second\cubed\ampere\squared\per\kilogram\per\meter\squared}$.} obtaining (here exemplified with the mid-point formula)
\begin{align}
	\bar \sigma &= \frac{\mathrm{e}^2}{\hbar}\, \frac{\Nt^2 \dt^2}{\pi}\, G(\beta/2)\,.\label{eq:sigma_bar}
\end{align}

\begin{figure*}
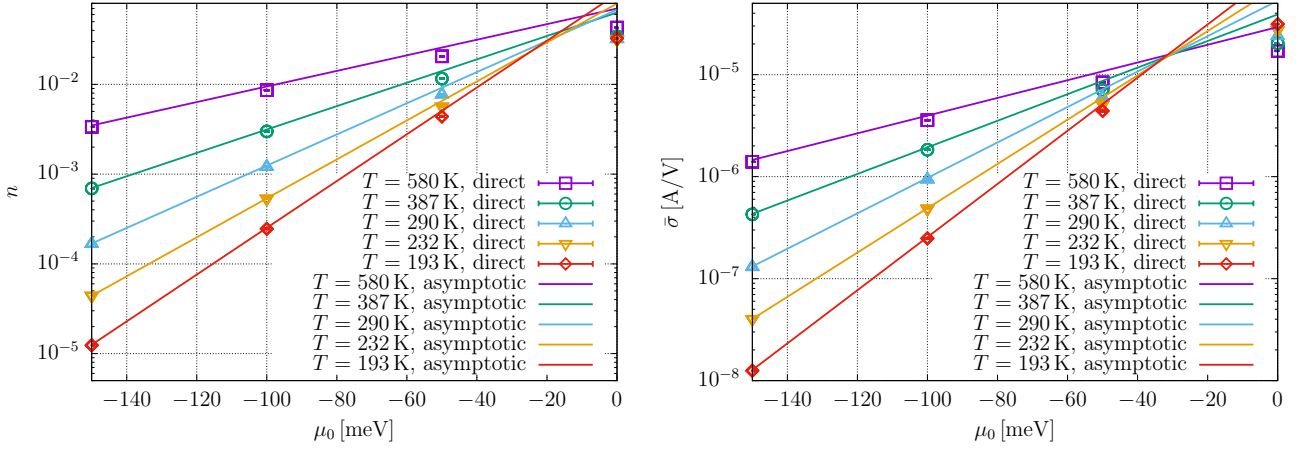

	\centering
	\resizebox{0.98\textwidth}{!}{{\large\input{data/n_avg}\input{data/cond}}}
	\caption{Average charge density $\erwartung{n}$ (left) and conductivity $\sigma$ (right) as a function of the effective chemical potential $\mu_0$ for different temperatures $T=\nicefrac 1\beta$. Data points have been obtained by `direct' numerical simulations at the respective chemical potential, while lines show the `asymptotic' behaviour for $\mu_0\rightarrow-\infty$ and have been calculated directly in the single-particle/low-density limit.}
	\label{fig:el_num-cond-supl}
\end{figure*}

\begin{figure*}
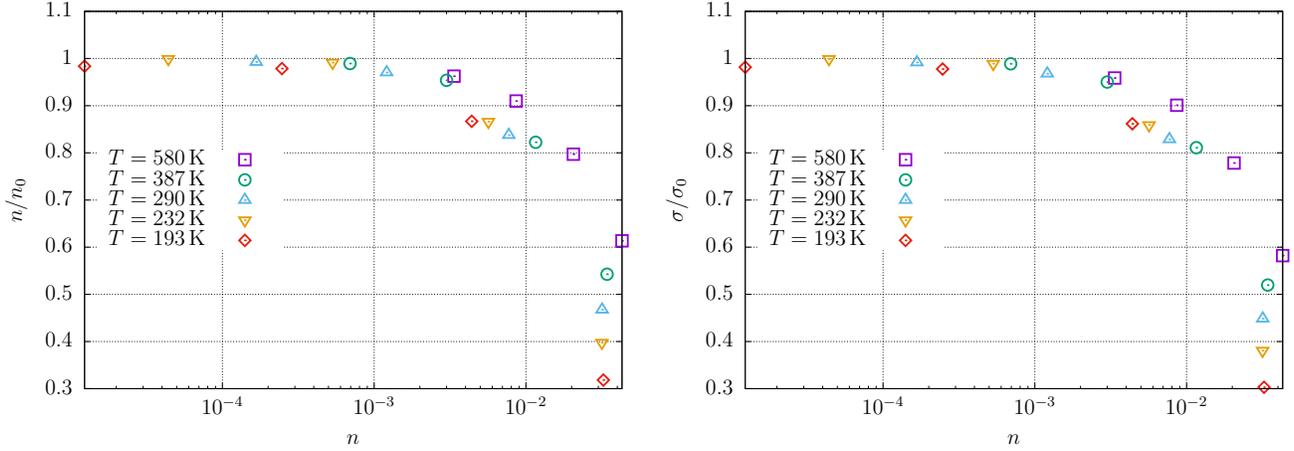

	\centering
	\resizebox{0.98\textwidth}{!}{{\large\input{data/n_avg_rel}\input{data/cond_rel}}}
	\caption{Relative charge density $\erwartung{n}$ (left) and electrical conductivity $\sigma$ (right) normalised by the asymptotic zero-filling value $\erwartung{n}_0$ and $\sigma_0$, respectively, as a function of the electron number $\erwartung{n}$ (calculated directly at finite filling) for different temperatures $T$. Error bars are of similar size as the data points and have been omitted for clarity.}
	\label{fig:el_num-cond-rel-supl}
\end{figure*}

We show the results of both, charge density and conductivity, side by side in Fig.~\ref{fig:el_num-cond-supl}. In addition, Fig.~\ref{fig:el_num-cond-rel-supl} shows their relative deviation from the asymptotic prediction in the low-density limit (similar to Fig.~\ref{fig:el_num-cond} of the \manuscript). 

\subsection{Charge Mobility}\label{sec:mobility}

Finally, the electron mobility is simply the conductivity per charge
\begin{align}
	\mu &= \frac{\sigma}{\mathrm{e} \erwartung{n}}\,.
\end{align}
After plugging in all the natural constants, the relation reads (again, using the mid-point estimator (\ref{eq:sigma_bar}) as an example):
\begin{align}
	\mu &\approx \SI[per-mode=fraction]{1.519e15}{\per\volt\per\second} \cdot \frac{\Nt^2}{\pi} \cdot \frac{\tilde{G}(\beta/2)}{\erwartung{n}}\\
	&= \SI[per-mode=fraction]{0.1519}{\centi\meter\squared\per\volt\per\second} \cdot \frac{1}{\si{\angstrom\squared}} \cdot \frac{\Nt^2}{\pi} \cdot \frac{\tilde{G}(\beta/2)}{\erwartung{n}}\,.
\end{align}

\section{Simulations in the low-density limit}\label{sec:zero-filling}

In the physically relevant regime of very low charge density $\erwartung{n}\lesssim \num{0.01}$ it makes sense to consider charge carriers as a perturbation. Physically, the limit of exactly zero charge density $\erwartung{n} = 0$ corresponds to free phonons which can be simulated very easily, even with an exact quantum mechanical treatment. We can then calculate how an infinitesimal number of charge carriers scatters off the free phonon fields.

In this limit, the path integral weight in (\ref{eq:path_int_Q1}) is dominated by the phonon statistical weight $\eto{-S_\text{ph}\lrs{x\lr{\tau}}}$, which, upon discretization of imaginary time, is equivalent to a multivariate normal distribution of the imaginary-time phonon field variables $x_{k,a,t}$. Below we describe an efficient way to directly sample phonon variables from this multivariate distribution without invoking any Markov chain, which allows us to completely get rid of autocorrelations.

The probability distribution of rescaled phonon variables $y_t$ defined in (\ref{eq:rescaling}) can be read off from~\eqref{eq:hamilton_of_y} as 
\begin{align}
	y &\sim \eto{-\frac{1}{2\tilde \omega}\sum\limits_t\left(y_{t+1}-y_t\right)^2 -	\frac{\tilde\omega}{2} y^2}
  \label{eq:multivariate_normal}
\end{align}
As derived in Section~\ref{sec:fourier_acc}, this is equivalent to sampling in Fourier space from a normal distribution
\begin{align}
	Y_\xi &\sim \mathcal{N}\left(0, w_\xi^{-2}\right)
\end{align}
with inverse standard deviation $w_\xi$ as in equation~\eqref{eq:phonon_frequency}. Then the desired configuration is obtained via the inverse Fourier transformation $y=\mathcal{F}^{-1}[Y]$ over the time coordinate.

This sampling process has a runtime in $\ordnung{V \Nt\log \Nt}$ and no autocorrelation at all, making it faster than the HMC by many orders of magnitude. Indeed, the sampling process in this case is totally negligible as compared to the runtime required for fermionic measurements.

Fermionic observables naturally go to zero in the low-density limit $\mu_e \rightarrow -\infty$, so we have to rescale them in order to get meaningful results that can be extrapolated to small but non-zero $\erwartung{n}$. We define the effective fugacity as a rescaling parameter
\begin{align}
	\varepsilon &\coloneqq \eto{\beta\mu_0}\,,
\end{align}
which is exponentially small with negative effective chemical potential $\mu_0$, i.e.\ in the regime with a Fermi surface below the non-interacting band.

The rescaled charge density and current-current correlator are then well-defined and finite even in the limit of exactly zero density. They can be obtained in our numerical simulations calculating the simplified expressions
\begin{align}
	\lim\limits_{\varepsilon\rightarrow0}\frac{\erwartung{n}}{\varepsilon} &= \frac 1V \tr\left\{\prod_t M_t[\mu_0=0]\right\}\,,\label{eq:n_at_zero_filling}\\
	\begin{split}			
		\lim\limits_{\varepsilon\rightarrow0}\frac{G(\tau_t) }{\varepsilon}
		&= \frac{1}{2 \, \beta^2 \, V} \sum_{k,l}\left\langle
        \lr{d_a}_{\alpha} \, \lr{d_b}_{\alpha} \, \chi_{k,a,0} \chi_{l,b,t} 
        \left[
		\left(Q_{t}\right)_{l,k}\left(P_{t}\right)_{k+a,l+b}
		+\left(Q_{t}\right)_{l+b,k+a}\left(P_{t}\right)_{k,l}\right.\right.\\
		&\qquad\qquad\qquad\qquad\;\left.\left.\left.
		-\left(Q_{t}\right)_{l+b,k}\left(P_{t}\right)_{k+a,l}
		-\left(Q_{t}\right)_{l,k+a}\left(P_{t}\right)_{k,l+b}
		\right]\right\rangle\right|_{\mu_0=0}\,,
	\end{split}\label{eq:cur-cur_zero_filling}
\end{align}
which will be derived via Taylor expansion in $\varepsilon$ below. The averaging is now performed over a multi-variate normal distribution (\ref{eq:multivariate_normal}). These measurements have a runtime in $\ordnung{V^2 \, \Nt}$ as opposed to $\ordnung{V^3}+\ordnung{V^2 \, \Nt \log \Nt}$ required for the full HMC. Together with the absent autocorrelation this can easily reduce the runtime required to reach a given precision by an order of magnitude or more.

In the continuum limit, the expression (\ref{eq:cur-cur_zero_filling}) reduces to a simple expression 
\begin{equation}
\label{eq:current_current_low_gauss}
G_1\lr{\tau} =  
\vev{\tr 
 I_{\alpha}\lrs{x\lr{0}} 
 U\lr{0, \tau} 
 I_{\alpha}\lrs{x\lr{\tau}}
 U\lr{\tau, \beta}} ,
\end{equation}
where $I_{\alpha}$ is the single-particle current operator introduced in (\ref{eq:single_particle_current}), and averaging is performed over the multivariate Gaussian distribution of phonon field variables $x_{k,a,t}$ (Eq.~(\ref{eq:multivariate_normal}) in terms of the rescaled phonon variables $y$). The above expression can be easily obtained by a fugacity expansion similar to the one in Eqs.~(\ref{eq:det_expansion1}) and  (\ref{eq:det_expansion2}). The Monte Carlo estimator (\ref{eq:cur-cur_zero_filling}) and its continuum counterpart (\ref{eq:current_current_low_gauss}) correspond to the mobility $\mu = \sigma\lr{\omega \rightarrow 0}/n$ in the limit $\mu_e \rightarrow -\infty$, where both the charge density $n$ and the conductivity $\sigma\lr{\omega}$ tend to zero as $e^{\beta\mu_e}$, but their ratio remains finite.

In order to perform the derivation, we first rewrite the fermion matrix
\begin{align}
	M_t &= \eto{\tilde\mu_0}\eto{\left(\tilde\mu-\tilde\mu_0\right) + H(x_t)\dt}\\
	&\equiv \eto{\tilde\mu_0} M_t[\mu_0=0] \\
	\Rightarrow \prod_t M_t &= \eto{\beta\mu_0} \prod_t M_t[\mu_0=0]\\
	&= \varepsilon \prod_t M_t[\mu_0=0]\,,
\end{align}
which allows the Taylor expansion
\begin{align}
	\hat{M}^{-1} &= 1 - \varepsilon \prod_t M_t[\mu_0=0] + \ordnung{\varepsilon^2}\,.
\end{align}
Thus, to leading order in $\varepsilon$, we can replace $\hat{M}$ and $\hat{M}^{-1}$ by the identity everywhere, unless this replacement yields a result equal to zero. This formally justifies the free phonon approximation in the limit of low density as the fermion determinant reduces to $\det M = 1+\ordnung{\varepsilon}$ in this case.

Since the charge density vanishes in the low density limit, the leading order is linear in $\varepsilon$
\begin{align}
	\erwartung{n} &= 1 - \frac 1V \tr\left\{1 - \varepsilon \prod_t M_t[\mu_0=0] + \ordnung{\varepsilon^2}\right\}\\
	&= \varepsilon\frac 1V \tr\left\{\prod_t M_t[\mu_0=0]\right\} + \ordnung{\varepsilon^2}\\
	\Rightarrow \frac{\erwartung{n}}{\varepsilon} &= \frac 1V \tr\left\{\prod_t M_t[\mu_0=0]\right\} + \ordnung{\varepsilon}\,,
\end{align}
which proves formula~\eqref{eq:n_at_zero_filling}.

Similarly, expanding the constituent terms of the current-current correlator~\eqref{eq:cur-cur_corr} yields
\begin{align}
	\left(R_{t,0}\right)_{kl}\left(T_{0,t}\right)_{k'l'} &= \left(Q_{t}\hat M^{-1}\right)_{kl}\left(\hat M^{-1}P_{t}\right)_{k'l'}\\
	&= \left(\varepsilon^{1-\nicefrac{t}{\Nt}} Q_{t}[\mu_0=0]\left(1+\ordnung{\varepsilon}\right)\right)_{kl}\left(\left(1+\ordnung{\varepsilon}\right)\varepsilon^{\nicefrac{t}{\Nt}} P_{t}[\mu_0=0]\right)_{k'l'}\\
	&= \varepsilon \left( Q_{t}[\mu_0=0]\right)_{kl}\left( P_{t}[\mu_0=0]\right)_{k'l'} + \ordnung{\varepsilon^2}\\
	\Rightarrow \frac{1}{\varepsilon}\left(R_{t,0}\right)_{kl}\left(T_{0,t}\right)_{k'l'} &= \left.\left( Q_{t}\right)_{kl}\left( P_{t}\right)_{k'l'}\right|_{\mu_0=0} + \ordnung{\varepsilon}\,.
\end{align}
Plugging all four terms of this type into equation~\eqref{eq:cur-cur_corr}, leads to the result~\eqref{eq:cur-cur_zero_filling} in the low-density limit.

Calculating the current-current correlators by a simple averaging over a multivariate normal distribution as in (\ref{eq:cur-cur_zero_filling}) has the advantage of being technically very simple and completely free of autocorrelations, and provides a useful alternative to the Hybrid Monte-Carlo algorithm described in Subsection~\ref{subsec:single_particle_HMC} which samples phonon variables with the non-local weight $\tr U\lr{0, \beta} e^{-S_\text{ph}\lr{x_{\tau}}}$.

Since the factors $Q_t$ and $P_t$ in (\ref{eq:cur-cur_zero_filling}) contain terms that are exponential in phonon variables $x_{k,a,t}$, the resulting distribution of $G\lr{\tau_t}$ appears to be a heavy-tailed log-normal distribution \cite{Kaplan:1106.0073}. 

In contrast, when including the factor $\tr U\lr{0, \beta}$ in the probability weight of phonon variables, sampled by the HMC algorithm in Subsection~\ref{subsec:single_particle_HMC}, 
the estimator (\ref{eq:current_current_low_gauss}) becomes
\begin{equation}
\label{eq:current_current_low_improved}
G_1\lr{\tau} =  
\vev{\frac{\tr 
 I_{\alpha}\lrs{x\lr{0}} 
 U\lr{0, \tau} 
 I_{\alpha}\lrs{x\lr{\tau}}
 U\lr{\tau, \beta}}{2 \, \tr U\lr{0, \beta}}
} .
\end{equation}
There is a significant cancellation of statistical fluctuations between the observable (\ref{eq:current_current_low_gauss}) in the numerator and the factor $\tr U\lr{0, \beta}$ in the denominator of (\ref{eq:current_current_low_improved}), which results in significantly smaller statistical errors at a fixed number of statistically independent phonon field configurations used for Monte-Carlo averaging. In contrast to (\ref{eq:current_current_low_gauss}), this estimator does not have a heavy-tailed distribution. 

For this reason the more complex HMC algorithm of Subsection~\ref{subsec:single_particle_HMC} can still outperform the algorithm described in this Section in some range of Hamiltonian parameters and temperatures. For the purpose of the present work, however, both algorithms are extremely fast and reach permille-level statistical errors in $\mathcal{O}\lr{1}$ CPU-hours.

\nocite{WangPhysChemChemPhys2015BandRenormalization}
\nocite{TroisiNatureMaterials2017,VongJPhysChemLett2022,Mannsfeld:2011}
\nocite{Giannini:2303.13163}
\nocite{VongJPhysChemLett2022}
\nocite{TroisiNatureMaterials2017,Northrup:2011}
\nocite{TroisiNatureMaterials2017,VongJPhysChemLett2022}
\nocite{TroisiNatureMaterials2017,Ishii:2020}
\nocite{TroisiNatureMaterials2017}
\nocite{WangPhysChemChemPhys2015BandRenormalization,VongJPhysChemLett2022}
\nocite{TroisiNatureMaterials2017,FratiniNikolkaNatMat2020}
\nocite{WangPhysChemChemPhys2015BandRenormalization,VongJPhysChemLett2022}
\nocite{TroisiNatureMaterials2017,FratiniNikolkaNatMat2020}
\nocite{VongJPhysChemLett2022,TroisiNatureMaterials2017,NematiaramADFM2020}
\nocite{MontvayMuenster}
\nocite{trotter_original}
\nocite{Verlet:1967}
\nocite{Blankenbecler:PhysRevD.24.2278}
\nocite{Duane1987,Blankenbecler:PhysRevD.24.2278}
\nocite{lattice23}
\nocite{Duane1987,Assaad:1708.03661}
\nocite{Ostmeyer:2024amv}
\nocite{ssh_simulations}
\nocite{Scalettar:2203.01291}
\nocite{Scalettar:2203.01291}
\nocite{Scalettar:2203.01291}
\nocite{Scalettar:2203.01291}
\nocite{Scalettar:2203.01291}
\nocite{monte_carlo_errors}
\nocite{monte_carlo_errors}
\nocite{Verlet:1967,trotter_omelyan}
\nocite{OMELYAN2003272}
\nocite{Blankenbecler:PhysRevD.24.2278}
\nocite{Peierls:33:1,Buividovich:12:1}
\nocite{Tripolt:1801.10348}
\nocite{ulybyshev_code_2017}
\nocite{Buividovich:12:1}
\nocite{Kaplan:1106.0073}
\nocite{REVTEX42Control}
\nocite{apsrev42Control}

\end{document}